\documentclass[5p, twocolumn]{elsarticle} 

\usepackage{graphicx}
\usepackage{amsmath}
\usepackage{amssymb}
\usepackage{hyperref}
\hypersetup{
colorlinks,
citecolor=blue,
filecolor=blue,
linkcolor=blue,
urlcolor=blue}
\begin{document}

\title{Scaling of a Fast Fourier Transform and a Pseudo-spectral Fluid Solver up to 196608 cores}

\author[iitk]{Anando G. Chatterjee}
\ead{anandogc@iitk.ac.in}
\author[iitk]{Mahendra K. Verma }
\ead{mkv@iitk.ac.in}
\author[iitk]{Abhishek Kumar \corref{cor1}}
\ead{abhishek.kir@gmail.com}
\cortext[cor1]{Corresponding author}
\author[kaust]{Ravi Samtaney}
\ead{ravi.samtaney@kaust.edu.sa}
\author[kaustc]{ Bilel Hadri}
\ead{bilel.hadri@kaust.edu.sa}
\author[kaustc]{Rooh Khurram}
\ead{rooh.khurram@kaust.edu.sa}
\address[iitk]{Department of Physics, Indian Institute of Technology Kanpur, Kanpur 208016}
\address[kaust]{Mechanical Engineering, Division of Physical Science and Engineering, King Abdullah University of Science and Technology - Thuwal 23955-6900, Kingdom of Saudi Arabia}
\address[kaustc]{KAUST Supercomputing Laboratory, King Abdullah University of Science and Technology - Thuwal 23955-6900, Kingdom of Saudi Arabia}

\begin{abstract}
In this paper we present scaling results of a FFT library, FFTK, and a pseudospectral code, Tarang, on grid resolutions up to $8192^3$ grid using  65536 cores of Blue Gene/P and 196608 cores of Cray XC40 supercomputers.   We observe that communication dominates computation, more so on the Cray XC40.  The computation time scales as $T_\mathrm{comp} \sim p^{-1}$, and the communication time  as $T_\mathrm{comm} \sim n^{-\gamma_2}$ with $\gamma_2$ ranging from 0.7 to 0.9 for Blue Gene/P, and from 0.43 to 0.73 for Cray XC40.  FFTK, and the fluid and convection solvers of Tarang exhibit weak as well as strong scaling nearly up to 196608 cores of Cray XC40.  We perform a comparative study of the performance on the Blue Gene/P and Cray XC40 clusters.
\end{abstract}

\maketitle

\section{Introduction}
\label{sec:intro}

 The Fast Fourier Transform (FFT), first discovered by Cooley and Tukey~\cite{Cooley:AMS1965}, is an important tool for image and signal processing, and radio astronomy.  It is also used to solve partial differential equations, fluid flows, density functional theory, many-body theory, etc.   For a three-dimensional $N^{3}$ grid, FFT has large time complexity $\mathcal{O}(N^{3} \log N^3)$ for large $N$ (e.g. 4096 or 8192).  Hence, parallel algorithms have been devised to compute FFT of large grids.

\begin{figure}[htbp]
\begin{center}
\includegraphics[scale = 1]{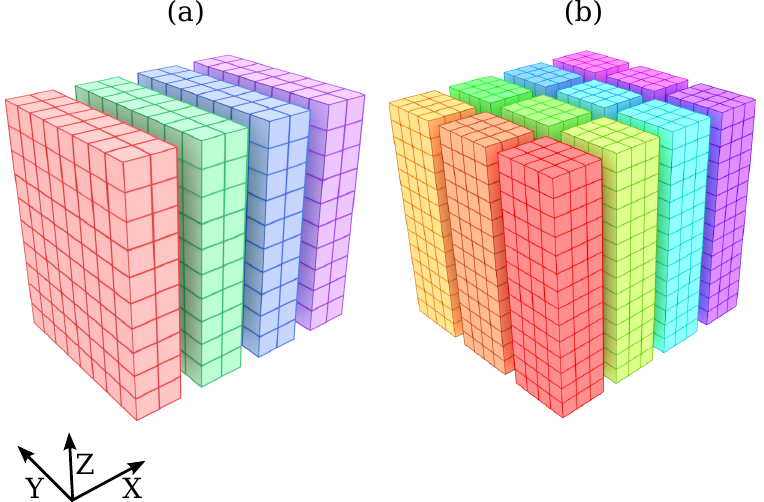}
\end{center}
\setlength{\abovecaptionskip}{0pt}
\caption{(a) Slab decomposition of an array. (b) Pencil decomposition of an array}
\label{fig:slab_pencil}
\end{figure}

One of the most popular opensource FFT libraries is FFTW (Fastest Fourier Transform in the West)~\cite{FFTW05,fftw:web} in which a three-dimensional (3D) array is divided into slabs (hence called {\em slab decomposition}) as shown in Fig.~\ref{fig:slab_pencil}(a).  Hence, we can employ a maximum of $N$ cores in FFTW operations on an array of size $N^3$. This is a severe limitation since present-day supercomputers offer several hundreds of thousands of cores for use.  To overcome this issue, Pekurovsky \cite{Pekurovsky:SC12} employed a {\em pencil decomposition} in which the data is divided into pencils, as shown in Fig.~\ref{fig:slab_pencil}(b).  This method allows usage of a maximum of $N^2$ cores, equal to  the maximum number of pencils.

In this paper we implement a pencil-based FFT  using the  algorithm of P3DFFT, and then construct a pseudo-spectral fluid solver.  As described earlier, the most well-known  pencil-based FFT library is P3DFFT, written by Pekurovsky~\cite{Pekurovsky:SC12} who reported that the total time $T$ for a FFT  operation is a sum of  computation time $a/p$ and  communication time $b/p^{2/3}$, where $p$ is the number of cores.  This scaling was deduced based on runs using a grid of  $8192^3$ points on a Cray XT5 with a 3D torus interconnect, and 65536 cores. In these tests, the  communication time dominates the computation time due to the \texttt{MPI\_Alltoall} data transfer.  Chan {\em et al.}~\cite{Chan:HPIC2008} studied scaling of P3DFFT on a 16384 nodes of Blue Gene/L system; they reported that a combination of the network topology and the communication pattern of P3DFFT can affect performance. 

Pippig and Potts~\cite{Pippig:2010wi} devised a similar FFT named  PFFT, and ran it on  a large number of cores; they observed that PFFT has a similar scaling as P3DFFT. {Richards {\em et al.}~\cite{Richards:SC2009} performed scalability analysis for their two-dimensional pencil FFT library on Blue Gene/P.} Czechowski {\em et al.}~\cite{Czechowski:ACM2012} analysed the memory hierarchy traffic and network communication in GPU-based FFT,  DiGPUFFT.  Mininni {\em et al.}~\cite{Mininni:PC2011} employed hybrid scheme (MPI + OPENMP) to use large number of cores optimally; their FFT implementations scales well on  $1536^3$ and  $3072^3$ grids for approximately 20000 cores with 6 and 12 threads on each socket.

We have devised another pencil-based FFT called FFTK (FFT Kanpur) and tested it on 65536 cores of Blue Gene/P (Shaheen I)  and 196608 cores of Cray XC40 (Shaheen II) of KAUST.    We computed separately the time required for the computation and communications during the FFT process; we showed that the computation time scales linearly, while the communication component approaches the ideal bisection bandwidth scaling for large arrays.  In this paper we compare the performance of FFTK on Blue Gene/P and Cray XC40, and show that the relative speed of cores and switch matters for the efficiency.   We show later that the per-core efficiency of Cray XC40 is lower than that of Blue Gene/P because the speed of the interconnect of  Cray XC40 has not increased in commensurate with the speed of the processor.  FFTK library is available for download at {\em http://www.turbulencehub.org/codes/fftk}.

FFT is used extensively in a pseudo-spectral method, which is one of the most accurate methods for solving differential equations due to the exponential convergence of derivative computations in this method~\cite{Boyd:Book,Canuto:SpectralFluid_book}. In addition, a major advantage of the spectral method is that it allows for a convenient scale-by-scale analysis of the relevant quantities.  Consequently, we can compute   interactions and energy transfers among structures at different scales using the spectral method.  Such scale-by-scale analysis is generally quite cumbersome in finite-difference and finite-volume solvers.   Note that the convolution stemming from nonlinear terms in a partial differential equation is computed using the FFT.  Though spectral method can be used to solve various types of partial differential equations (PDEs), in this paper we focus on the PDEs for fluid flows.

Turbulence remains one of the unsolved problems of classical physics, and no analytical solution of fluid equations in the turbulent limit is  available at present.  Hence numerical simulation is very handy for the analysis of turbulent flows.  Unfortunately, the grid resolution and computational time required for turbulence simulations is very large. Hence such simulations are performed on large high performance computing clusters (HPC).

Many researchers~\cite{Yokokawa:PS2002,Kaneda:PF2003,Donzis:2008Conf,Yeung:PNAS2015,Rosenberg:PF2015,Donzis:PF2008,Donzis:FTC2010,Donzis:JFM2010a,Yeung:PF2005,Yeung:JFM2013,Rorai:PRE2015,Dallas:PRL2015}  have performed high resolution turbulence simulations.  Yokokawa {\em et al.}~\cite{Yokokawa:PS2002} performed first turbulence simulation on $4096^3$ grid using the {\em Earth Simulator}.  Donzis {\em et al.}~\cite{Donzis:2008Conf} performed turbulence simulation on $4096^3$ grid using P3DFFT library;  they employed 32768 cores of Blue Gene/L and Cray XT4, and  reported  that the effective performance of the FFT is  approximately 5\% of the peak performance due to the extensive communication and cache misses.  Yeung {\em et al.}~\cite{Yeung:PNAS2015} performed pseudo spectral simulations of  fluid turbulence on one of the highest resolution grids ($8192^3$) to study extreme events.  Rosenberg {\em et al.}~\cite{Rosenberg:PF2015} simulated rotating stratified turbulence on a $4096^3$ grid and studied its energy spectrum.

We have implemented a pseudospectral code named   Tarang based on the FFTK library.  Tarang is a  parallel and C++ code written as a general PDE (partial-differential equation) solver.  Using Tarang, we can compute incompressible flows involving pure fluid, magnetohydrodynamic flows~\cite{Kumar:EPL2014}, liquid metal flows~\cite{Reddy:PF2014},  Rayleigh-B\'{e}nard convection~\cite{Verma:PRE2012,Pandey:PRE2014,Kumar:PRE2014}, rotating convection~\cite{Pharasi:PF2013}, and rotating flows~\cite{Sharma:preprint2016},  etc.  Simulation of convective flows involves rigid walls for which periodic boundary condition is not applicable.  For such simulations, no-slip or free-slip boundary conditions are employed.  Another major feature of Tarang is its rich library for computing the energy transfers in turbulence, e.g., energy flux, shell-to-shell and ring-to-ring energy transfers, etc.~\cite{Kumar:PRE2014,Verma:PR2004,Nath:2016}.  Tarang also enables us to probe any point in the Fourier space or in the real space~\cite{Reddy:PF2014}. In the present paper we empirically demonstrate that the flow solvers of Tarang scale  well on HPC supercomputers.

The outline of the paper as follows: In Sections~\ref{sec:Num_Sch} and~\ref{sec:parallel}, we describe the numerical scheme and parallelization strategy.  Section~\ref{sec:hpc} contains a brief discussion about the HPC systems used for scaling analysis.  We describe our scaling results for FFTK, the fluid solver, and the Rayleigh-B\'{e}nard solver in Sections~\ref{sec:fft_results},~\ref{sec:result_fluid}, and~\ref{sec:result_rbc}, respectively.   We run  FFTK on Blue Gene/P and benchmark this against the performance of P3DFFT on the same platform. We do not undertake a detailed comparison between the two FFT packages due to lack of space and comparison resources.  We also deduce the Kolmogorov-like spectrum for Rayleigh-B\'{e}nard convection using analysis of data from a  $4096^3$ grid simulation.
We conclude in Section~\ref{sec:conclusions}.

\section{Numerical Scheme}
\label{sec:Num_Sch}
In a pseudospectral code, approximately 70\% to 80\% of the total time is spent on the forward and inverse Fourier transforms.  In this section we briefly explain the numerical schemes for FFT and the spectral solver Tarang.

\subsection{Fast Fourier Transform}
\label{sec:fft}
The inverse Fourier transform  is defined as
\begin{equation}
f(x,y,z) = \sum_{k_x, k_y, k_z} \hat{f}(k_x, k_y, k_z) \phi_{k_x}(x) \phi_{k_y}(y)  \phi_{k_z}(z), 
\end{equation}
where $\hat{f}(k_x, k_y, k_z)$ is the Fourier transform of $f(x,y,z)$. Here we compute $f(x,y,z)$ from $\hat{f}(k_x, k_y, k_z)$.  The functions $\phi_k(s)$ are the basis functions that appear in the following forms:
\begin{subequations}
\begin{eqnarray}
\mathrm{Fourier}: & \phi_k(s) = & \exp(i k s), \\
\text{Sine}: & \phi_k(s) = & 2 \sin( k s), \\
\text{Cosine}: & \phi_k(s) = & 2 \cos(k s),
\end{eqnarray}
\end{subequations}
where $k$ could be $k_x$, $k_y$, or $k_z$, and $s$ could be $x$, $y$, or $z$.  We use Fourier basis function for the periodic boundary condition, and employ the sine and cosine basis functions for the free-slip boundary condition. The FFTK library can perform the above transformations in different combinations. For example, for the periodic boundary condition along the three directions, we employ Fourier basis function
\begin{equation}
\exp(i k_x x +i k_y y + i k_z z).
\end{equation}
But for the free-slip boundary condition at all the three walls,  $u_z$, the $z$-component of the velocity  is expanded using the basis function
\begin{equation}
8 \cos(k_x x) \cos(k_y y) \sin(k_z z).
\end{equation}
Similar schemes are used for $u_x$ and $u_y$.   We term the above two basis functions as \texttt{FFF} and \texttt{SSS} respectively.  In a similar fashion, we  define other basis functions---\texttt{SFF} for one free-slip wall direction and two periodic directions,  \texttt{SSF} for two free-slip wall directions and one periodic direction, and \texttt{ChFF} for one no-slip wall direction and two periodic directions.  Note that the inverse of sine (or cosine) basis function is sine (or cosine) itself, but $\exp(i k x)$ and $\exp(-i k x)$ are mutual inverses of each other. 

In Sec.~\ref{sec:parallel} we will describe the FFT implementation  in our library.

\subsection{Spectral Solver}
\label{sec:solver}

As described in the introduction, the pseudospectral scheme is one of the most accurate methods to solve partial differential equations.  In the following, we describe its implementation for computing fluid flows.  The incompressible fluid equation is  described using the celebrated Navier Stokes equation, which is
\begin{eqnarray}
\frac{\partial{\textbf{u}}}{\partial{t}} + ({\bf u}\cdot\nabla){\bf u} &=& -\frac{1}{\rho}\nabla{p}+ \nu\nabla^2 {\bf u}, \label{eq:NS}\\
\nabla \cdot {\bf u} & = & 0, \label{eq:continuity1}
\end{eqnarray}
where ${\bf u}$ is the velocity field, $p$ is the pressure field, and $\nu$ is the kinematic viscosity.  For simplicity we take the density of the fluid, $\rho$, to be unity.  We rewrite the above equations in Fourier space:
\begin{eqnarray}
(\partial_{t} + \nu k^2)\hat{u}_j ({\bf k}, t)  & = & -i k_l \widehat{u_l u_j}({\bf k},t)   -  i k_j \hat{p} ({\bf k},t), \label{eq:NSk}\\
  k_j \hat{u}_j ({\bf k})   & = & 0, \label{eq:continuity2}
\end{eqnarray}
where $i = \sqrt{-1}$. The above equations are time advanced using standard methods, e.g., Runge Kutta scheme.  The $\widehat{u_l u_j}$ term of Eq.~(\ref{eq:NSk}) becomes a convolution in spectral space that is very expensive to compute. Orszag~\cite{Orszag:SAPM1972} devised an efficient scheme in which $\widehat{\bf u}({\bf k},t)$ is transformed to real space, components of which are multiplied with each other, and the product is then transformed back to Fourier space.   Due to the multiplication of arrays in real space, this method is called  {\em pseudospectral method}.  This multiplication however generates aliasing errors, which are mitigated by filling up only 2/3 of the array in each direction.  See Boyd~\cite{Boyd:Book} and Canuto~\cite{Canuto:SpectralFluid_book} for details. 

A spectral transform is  general, and it can involve basis functions from Fourier series, sines and cosines, Chebyshev polynomials, spherical harmonics, or a combination of these functions. The FFTK library uses Fourier, sines, and cosine functions only. We plan to incorporate Chebyshev polynomials and spherical harmonics in the future.   In this paper, we solve Eqs.~(\ref{eq:NSk},\ref{eq:continuity2}) in a periodic box (\texttt{FFF} basis) using Tarang. Therefore we illustrate the implementation and usage of \texttt{\texttt{FFF}} basis, and then describe scaling analysis of FFT, and the fluid and convection solvers.

We have computed fluid and magnetohydrodynamic flows, liquid metal flows,  Rayleigh-B\'{e}nard convection (RBC), rotating convection, and rotating flows using Tarang.  In the following we will illustrate one of the above modules, the RBC solver, whose governing equations are 
\begin{eqnarray}
\frac{\partial{\textbf{u}}}{\partial{t}} + (\mathbf{u} \cdot \nabla)\mathbf{u} & = & -\nabla \sigma + \alpha g \theta \hat{z} + \nu{\nabla}^2 \mathbf{u}, \label{eq:u} \\
\frac{\partial{\theta}}{\partial{t}} + (\mathbf{u} \cdot \nabla)\theta & = & \frac{\Delta}{d} u_{z} + \kappa{\nabla}^{2}\theta, \label{eq:th} \\
\nabla \cdot {\bf u} & = & 0, \label{eq:incompressible}
\end{eqnarray}
where $\theta$ is   the temperature fluctuation from the steady conduction state,
\begin{equation}
T(x,y,z) = T_c(z) + \theta(x,y,z),
\end{equation}
with $T_c$ as the conduction temperature profile), $\sigma$ is the pressure fluctuation, $\hat{z}$ is the buoyancy direction, $\Delta$ is the temperature difference between the two plates that are separated by a distance $d$,  $\nu$ is the kinematic viscosity, and $\kappa$ is the thermal diffusivity.  

We solve a nondimensionalized version of the RBC equations, which are obtained using $d$ as the length scale, $(\alpha g \Delta d)^{1/2}$ as the velocity scale, and $\Delta$ as the temperature scale:
\begin{eqnarray}
\frac{\partial{\textbf{u}}}{\partial{t}}+ (\textbf{u}\cdot \nabla)\textbf{u} & = & -\nabla\sigma +  \theta \hat{z}
+   \sqrt{\frac{\mathrm{Pr}}{\mathrm{Ra}}}\nabla^{2}\textbf{u},\label{eq:u_Usmall_Plarge}\\
\frac{\partial{\theta}}{\partial{t}}+(\textbf{u}\cdot\nabla)\theta  & = & u_{z} + \frac{1}{\sqrt{\mathrm{Ra Pr}}} \nabla^{2}\theta.\label{eq:T_Usmall_Plarge}
\end{eqnarray}
Here the two important nondimensional parameters are the Rayleigh number $\mathrm{Ra}=\alpha g \Delta d^3/\nu \kappa$, and the Prandtl number $\mathrm{Pr}=\nu/\kappa$.  We perform our simulations in a cubic fluid domain of unit size in each direction.

For the scaling analysis, we solve the RBC equations (Eqs.~(\ref{eq:u_Usmall_Plarge}, \ref{eq:T_Usmall_Plarge})) with \texttt{FFF} and  \texttt{SFF} basis functions.  We also perform a production run for computing the energy spectrum and energy flux during the statistical steady state.  Note that the  \texttt{SFF} basis function corresponds to the free-slip boundary condition for which  the velocity field at the top and bottom plates ($z=0,1$):
\begin{equation}
u_z= \partial_z {u_x}=\partial_z {u_y}  =  0, 
\end{equation}
and periodic boundary conditions imposed on the vertical side walls.  For the temperature field, we employ  isothermal boundary condition ($\theta=0$) at the top and bottom plates, and periodic boundary conditions at the side walls. 

\begin{figure*}[htbp]
\begin{center}
\includegraphics[scale = 0.9]{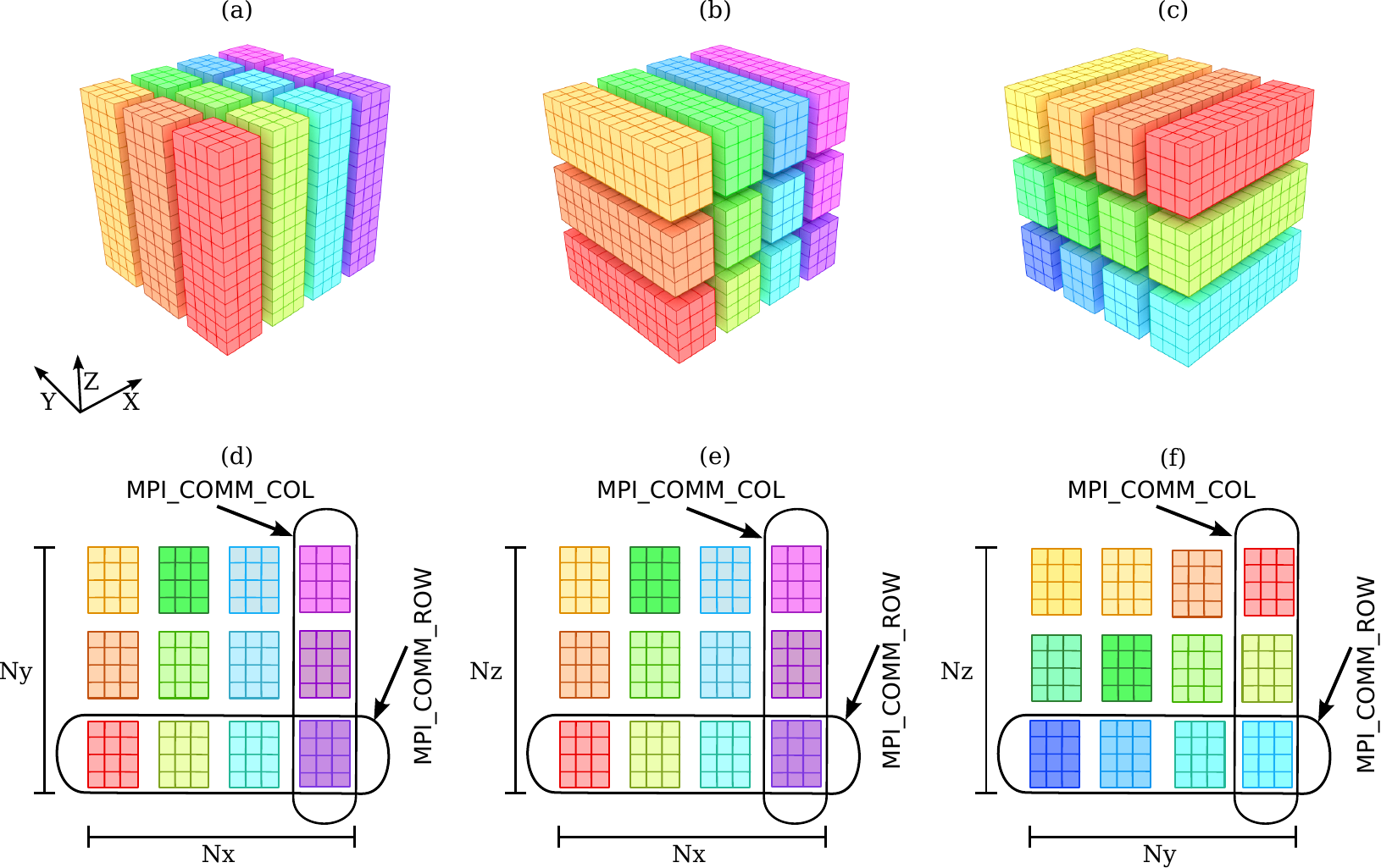}
\end{center}
\setlength{\abovecaptionskip}{0pt}
\caption{Pencil decomposition:  (a) real space data, (b) intermediate configuration, (c) data in Fourier space.  (d, e, f) Division of cores into $p_\mathrm{row}$ and $p_\mathrm{col}$ such that $p = p_\mathrm{row} \times p_\mathrm{col}$  as seen in $XY$, $XZ$, and $YZ$ projections respectively.  Here $N_x = N_y =  N_z = 12$.  In the subfigures (a,d), $p_\mathrm{row} = 3$, $p_\mathrm{col} =  4$, thus each core contains $N_{x}/ p_\mathrm{col} \times N_{y}/ p_\mathrm{row}  \times N_{z} = 3\times 4 \times 12$ data points. }
\label{fig:pencil_combined}
\end{figure*}

 Many fluid flow simulations, especially Rayleigh-B\'{e}nard convection, employ no-slip boundary condition that requires Chebyshev basis functions (\texttt{ChFF}), which are somewhat  complex to implement.   Note however that the \texttt{ChFF} basis  has its own limitations.  For example, the energy spectrum and flux computations performed in the Fourier space requires a uniform grid.  The collocation points in \texttt{ChFF} are nonuniform, and hence we need to interpolate the data to a uniform mesh that induces errors.  Therefore, the free-slip basis functions that involves uniform mesh are convenient for studying the energy spectrum and flux.  We remark the no-slip boundary condition captures the boundary layers near the walls.  The flow near the walls contributes to the energy spectrum at small scales or large wavenumbers, therefore the boundary layers at the top and bottom walls do not significantly impact the inertial-range energy spectrum and flux (see Sec. 7).  For such studies, the free-slip basis  suffices.

The above example illustrates that we can easily perform simulations with different boundary conditions using Tarang solvers just by changing the basis function in the input file.  We report our results for the RBC solvers in Sec.~\ref{sec:result_rbc}.

\section{Parallelization Strategy}
\label{sec:parallel}

In the following discussion we illustrate our implementation for the \texttt{FFF} basis.   We  divide the data equally among all the cores for load balancing.  For the pencil decomposition, we divide the data into rows and columns, and $p$ cores into a core grid $p_\mathrm{row} \times p_\mathrm{col} =  p$  as shown in Fig.~\ref{fig:pencil_combined}. The cores are divided into two MPI communicators: \texttt{MPI\_COM\_ROW} and \texttt{MPI\_COM\_COL} (see Fig.~\ref{fig:pencil_combined}). In real space [Fig.~\ref{fig:pencil_combined}(a)] each core has $N_x/p_\mathrm{col} \times N_y/ p_\mathrm{row} \times$ $N_z$ data points.

The forward FFT transform (from real to complex) follows the following set of steps: 
\begin{enumerate}
\item We perform forward FFT, \texttt{r2c} real-to-complex, along the Z-axis for each data column.
\item  We perform \texttt{MPI\_Alltoall} operation among the cores in \texttt{MPI\_COM\_COL} to transform the data from the real configuration [Fig.~\ref{fig:pencil_combined}(a)] to the intermediate configuration [Fig.~\ref{fig:pencil_combined}(b)]. 
\item After interprocess communication, we perform forward \texttt{c2c} (complex-to-complex) transform along the Y-axis for each pencil of the array. 
\item  We perform \texttt{MPI\_Alltoall} operation among the cores in \texttt{MPI\_COM\_ROW} to transform the data from the intermediate configuration [Fig.~\ref{fig:pencil_combined}(b)] to the Fourier configuration  [Fig.~\ref{fig:pencil_combined}(c)]. 
\item  Now we perform forward \texttt{c2c} transform along the X-axis for each  pencil [see Fig.~\ref{fig:pencil_combined}(c)].
\end{enumerate}
For  one-dimensional FFT operations, we use the FFTW transforms.  {During steps 2,4 of the above,  we employ transpose-free data transfer among the communicators, as described in~\ref{sec:appA}.  This scheme avoids local transpose during these processes.  However,  after the \texttt{MPI\_Alltoall}, the data along a column are not contagious. Hence, we need to employ strided FFT, which is efficient, and is provided in the FFTW library~\cite{fftw:web}.  Note, however, this is prone to cache misses since the columnar/row data are not contagious.  
See~\ref{sec:appA} for details.} We also remark that FFTK and Tarang use a meta-template C++ library, \texttt{Blitz}++~\cite{blitz,Blitz:paper}  for array manipulation; this library provides efficient operations for arrays. 



This completes the forward transform. The inverse transform is reverse of the above operation.  Note that the above strategy is general, and it works for all the basis functions.  Our library also works for two-dimensional (2D) data, for which we set $N_y=1$.  The intermediate state is avoided for 2D Fourier transforms.  Also note that the slab FFT can be performed by setting $p_\mathrm{row} = 1$, and  again,   the configuration (b) of Fig.~\ref{fig:pencil_combined} is avoided.

The other functions of a spectral solver that require parallelization are multiplication of arrays elements and input/output (I/O).  A multiplication of array elements is trivial to parallelize.  Since the data-size involved in high-resolution turbulence simulation is very large,  of the order of several terabytes, it is more efficient to use parallel I/O.  In our spectral code, we use  the \texttt{HDF5} library to perform parallel I/O.   

Our code has  important sets of functions to compute energy flux, shell-to-shell energy transfers, and ring-to-ring energy transfers.  These quantities are computed using FFT~\cite{Verma:PR2004,Nath:2016}.  For brevity, we omit discussion on these implementations in the present paper.  In Sec.~\ref{sec:result_rbc}, we will briefly describe the computation of the energy flux for RBC.    We have exploited this feature to implement \texttt{FFF},  \texttt{SFF},  \texttt{SSF},  \texttt{SSS} basis functions in two and three dimensions in a single code.   We also make use of efficient libraries such as \texttt{Blitz}++ and \texttt{HDF5}.  These are some of the unique features of FFTK and Tarang.


\section{About the HPC systems}
\label{sec:hpc}
 We performed  scaling tests of our FFT library and pseudospectral code  on Shaheen I, a Blue Gene/P supercomputer, and Shaheen II, a Cray XC40 supercomputer, of  King Abdullah University of Science and Technology (KAUST).  First we provide some details of these systems.

\subsection{Blue Gene/P}
The Blue Gene/P supercomputer consists of 16 racks with each rack containing 1024 quad-core, 32-bit, 850 MHz PowerPC compute nodes.  Hence the total number of cores in the system is 65536.  It also has  65536 GB of RAM.  The Blue Gene/P nodes are interconnected by a three-dimensional point-to-point  {\em torus} network.    The theoretical peak  speed of the Blue Gene/P supercomputer is 222 Tera FLOP/s (Floating point operations per second).

 \subsection{Cray XC40}

The Cray XC40 supercomputer  has 6174 dual-socket compute nodes each containing two Intel Haswell processors with 16 cores, with each core running at a clock speed of  2.3 GHz.  In aggregate, the system has  a total of 197568 cores and 790 TB of memory.  The compute nodes, contained in 36 water-cooled XC40 cabinets, are connected via the Aries High Speed Network.  Cray XC40 adopts a dragonfly topology that yields 57\% of the maximum global bandwidth between the 18 groups of two cabinets~\cite{Bilel:shaeen2}.   Shaheen II delivers a theoretical peak performance of 7.2 Peta FLOP/s and a sustained LINPACK performance of 5.53 Peta FLOP/s.

A parallel program involves computation time $T_\mathrm{comp}$ and communication time across nodes $T_\mathrm{comm}$~\cite{Sutou:ISHPC2003,Buchanan:book}.  Thus the total time $T$ for a parallel program is
\begin{equation}
T = T_\mathrm{comp} +  T_\mathrm{comm}.
\end{equation}
We report $T_\mathrm{comp}$ and $T_\mathrm{comm}$ for the execution of the FFTK library on the Blue Gene/P and Cray XC40 supercomputers.  Since the data is divided equally among all the cores, we expect $T_\mathrm{comp} \sim p^{-1}$, where $p$ is the number of cores; we observe the above scaling in all our tests.  Note that FFT, which is a dominant  operation in a pseudospectral solver, involves \texttt{MPI\_Alltoall} communications. Hence communication time is the most dominant component of the total time.


For the fluid and RBC solvers, it is difficult  to disentangle  the computation and communication times since they involve many functions, hence we report only the total time for these solvers.  These results are presented in the following  sections.

\section{Scaling results of FFTK}
\label{sec:fft_results}

We perform FFTK forward and inverse transforms  several times (100 to 1000) for large   $N^3$ grids, and then present an average time taken for a pair of forward and inverse transforms. We compute $T_\mathrm{comp}$ and $T_\mathrm{comm}$ for various combinations of grid sizes and number of cores, and observe that 
\begin{subequations}
\label{eq:time_sub_division}
\begin{align}
	T_\mathrm{comp}  &= \frac{1}{c_1} p^{-\gamma_1}, \\
	T_\mathrm{comm}  &= \frac{1}{c_1} n^{-\gamma_2}, \\
	T^{-1}        &= \frac{1}{C} p^\gamma,
\end{align}
\end{subequations}
where $c_{1}$, $c_{2}$, $C$, $\gamma_{1}$, $\gamma_{2}$, and $\gamma$ are constants, $p$ is the number of cores, and $n$ is the number of nodes.  Hence the total time per FFT operation is 
\begin{equation}
\label{eq:T}
T = c_1 D \left(\frac{1}{p^{\gamma_1}}\right) + c_2 D\left(\frac{1}{n^{\gamma_2}}\right) = C\left(\frac{1}{p^\gamma}\right),
\end{equation}
where $D=N^3$ is the data size.  We measure  $T_{\rm comp}$ and $T_{\rm comm}$ by computing the time taken by the respective code-segments using the MPI function   \texttt{MPI\_Wtime}.  We record the time when the process enters and leaves the code segment, and then take their difference that yields $T_\mathrm{comp}$ and $T_\mathrm{comm}$. 

After the above general discussion, we now describe our results specific to the Blue Gene/P supercomputer.

\subsection{Scaling on Blue Gene/P}
\begin{figure}[htbp]
\begin{center}
\includegraphics[scale = 0.8]{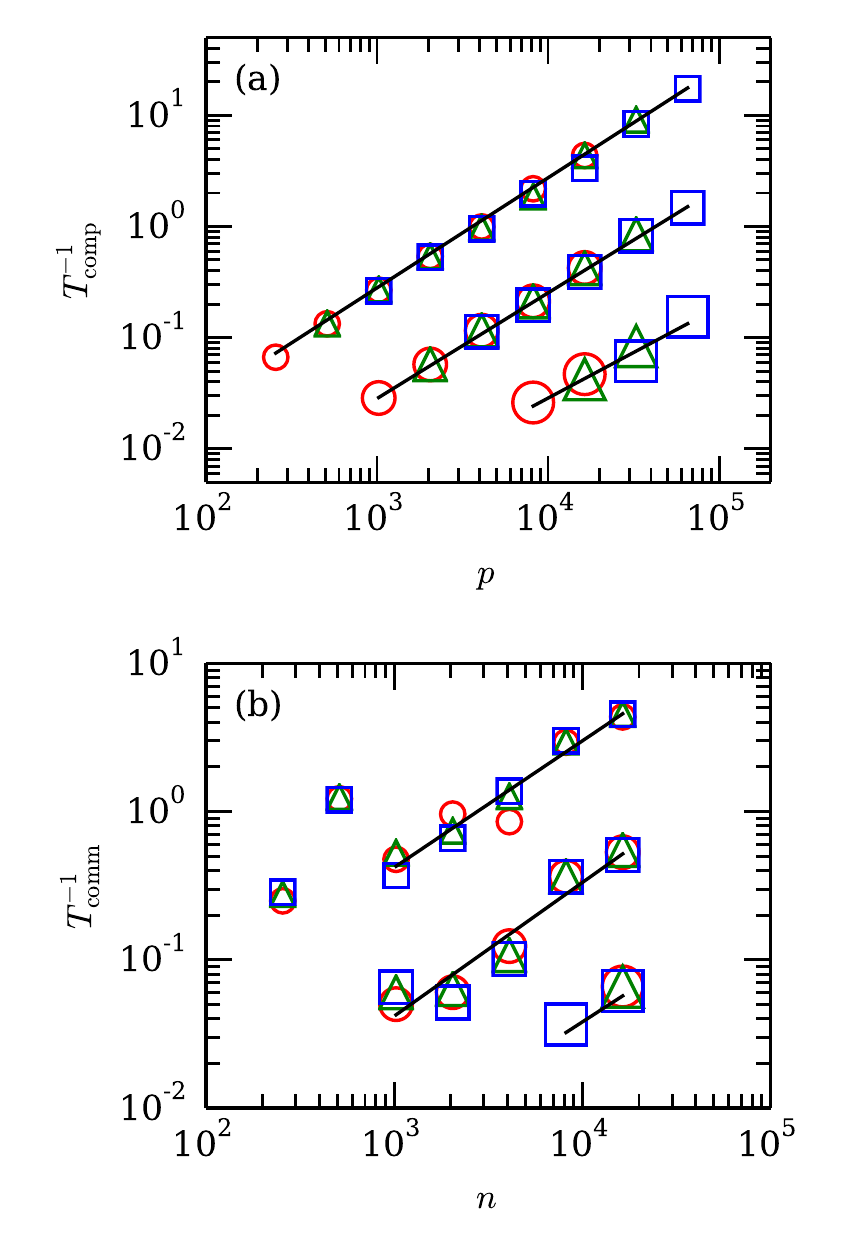}
\end{center}
\setlength{\abovecaptionskip}{0pt}
\caption{Scalings of the FFTK library on Blue Gene/P: (a) Plot of inverse computation time $T^{-1}_\mathrm{comp}$ vs.~$p$ (number of cores) for \texttt{1ppn}  (red circle),  \texttt{2ppn} (green triangle), \texttt{4ppn} (blue square).  Here \texttt{ppn} represents number of MPI processes per node. The data for  grids $2048^3$, $4096^3$, and $8192^3$ are represented by the same symbols but with increasing sizes.  The plots show that FFTK exhibits strong scaling in Blue Gene/P.  (b) Plot of inverse communication time $T^{-1}_\mathrm{comm}$ vs.~$n$ (number of nodes) with the above notation.  $T^{-1}_\mathrm{comm}$ for $p = 256$ and $512$  exhibits a better scaling  due to the slab decomposition employed. }
\label{fig:fftk_scaling_bg}
\end{figure}

 In Fig.~\ref{fig:fftk_scaling_bg} we plot $T_\mathrm{comp}$ and $T_\mathrm{comm}$.  Our results indicate that the computation time scales as $T_\mathrm{comp} \propto p^{-1}$.  However, the communication among the nodes takes maximum time in an FFT operation.  In the following discussion, we sketch the scaling arguments for $T_\mathrm{comm}$ that was first provided by Pekurovsky~\cite{Pekurovsky:SC12}.

A Blue Gene/P supercomputer has a torus interconnect for which we estimate the {\em bisection bandwidth} $B$, which is  defined as the available bandwidth when the network is bisected into two partitions.  For 3D torus, bisection bandwidth is proportional to the area in the network topology, hence $B \propto (n^\prime)^{2/3}$, where $n^\prime$ is the number of nodes used in communication.  

FFT involves \texttt{MPI\_Alltoall} communication, hence, in the slab division, $n' = n$, the total number of nodes, and the data to be communicated in the network is  $D = N^3$.  Therefore, the inverse of the communication time for each FFT is
\begin{equation}
T_\mathrm{comm,slab}^{-1} \sim \frac{B}{D} \sim n^{2/3}. 
\end{equation}
We also remark that the communication time depends on number of interacting nodes, not cores.  The inter-core or intra-node communication (among the cores within a node) is typically much faster than the inter-node communication across an interconnect.

\begin{figure}
\begin{center}
\includegraphics[scale = 0.8]{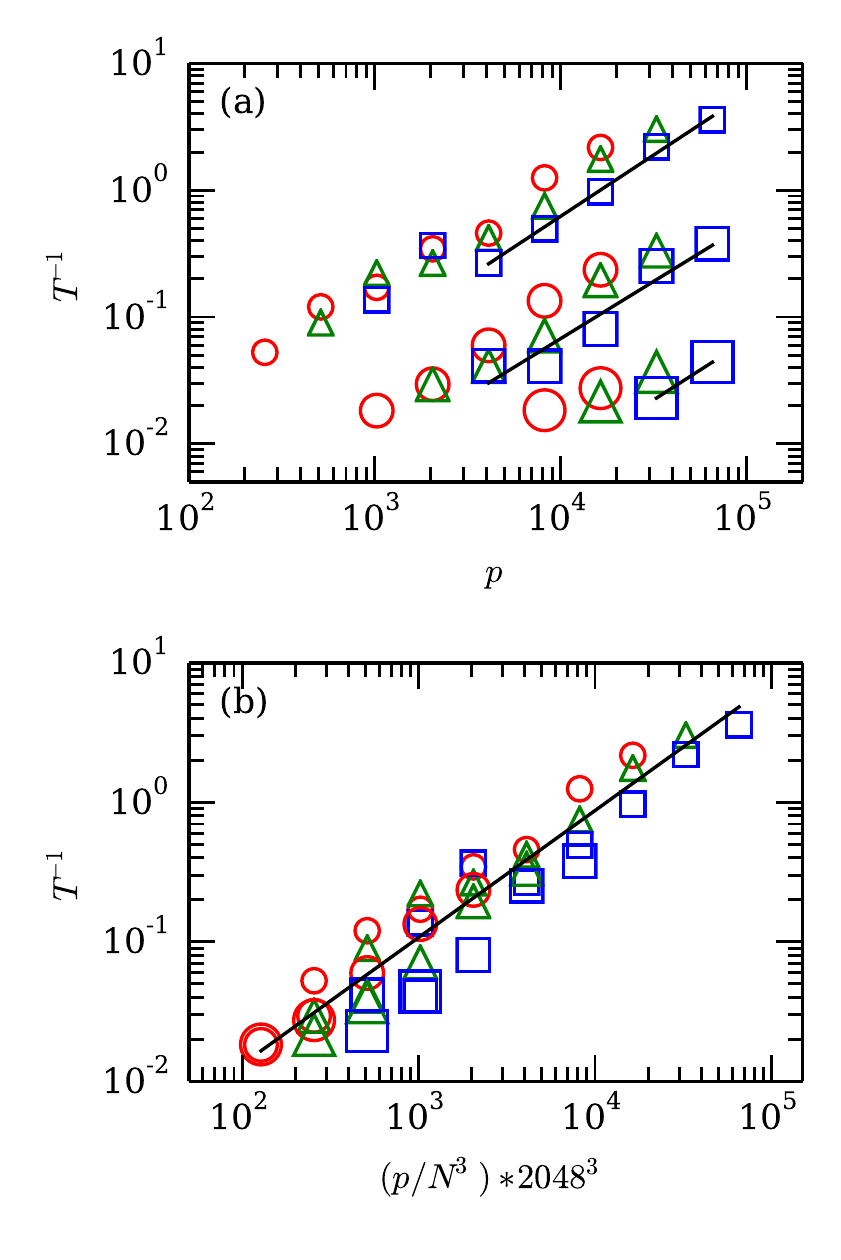}
\end{center}
\caption{Scalings of FFTK on Blue Gene/P: (a) Plot of  inverse of total time $T^{-1}$ vs.~$p$ exhibits strong scaling. (b) Plot of $T^{-1}$ vs.~$p/N^3$ exhibits weak scaling with the exponent  $\gamma=0.91\pm0.04$.  We follow the same colour and symbol convention as Fig. 3.}
\label{fig:fftk_scaling_bg_2}
\end{figure}

\setlength{\tabcolsep}{15pt}
\begin{table}[htbp]
\begin{center}
\caption{FFTK scaling on Blue Gene/P:   The exponents $\gamma_1$ for the computation time ($T_\mathrm{comp}$), $\gamma_2$ for the communication time  ($T_\mathrm{comm}$), and $\gamma$ for the total time  ($T$) [refer to Eq.~(\ref{eq:time_sub_division}) for definition]. The maximum nodes used is 16384  with  \texttt{1ppn}, \texttt{2ppn}, and \texttt{4ppn}.}
\hspace{20mm}
\begin{tabular}{c c c c }
\hline \hline \\[0.2 pt]
$\gamma$ & \texttt{ppn} &  $2048^3$ &   $4096^3$ \\[0.2 pt]
\hline \\[0.2 pt]
$\gamma_1$ & $1$ & $1.00 \pm 0.01$ & $0.97 \pm 0.01$  \\
           & $2$ & $1.00 \pm 0.02$ & $0.96 \pm 0.01$  \\
           & $4$ & $1.00 \pm 0.03$ & $0.95 \pm 0.03$  \\
\hline \\[0.2 pt]
$\gamma_2$ & $1$ & $0.7  \pm 0.1 $ & $0.9  \pm 0.1 $ \\
           & $2$ & $0.7  \pm 0.1 $ & $0.8  \pm 0.2 $ \\
           & $4$ & $0.7  \pm 0.1 $ & $0.8  \pm 0.2 $ \\
\hline \\[0.2 pt]
$\gamma$        & $1$ & $0.87 \pm 0.05$ & $0.94 \pm 0.05$ \\
           & $2$ & $0.81 \pm 0.05$ & $0.96 \pm 0.09$ \\
           & $4$ & $0.76 \pm 0.07$ & $0.9  \pm 0.1 $ \\
           \hline
\end{tabular}
\label{tab:fftk_exponents_bg}
\end{center}
\end{table}

In the pencil division, the nodes are divided into row nodes ($n_\mathrm{row}$)  and column nodes ($n_\mathrm{col}$).   We estimate  $n_\mathrm{row} \approx n_\mathrm{col} \approx n^{1/2}$.  Hence each communication within a row (or a column) involves $n' \approx n^{1/2}$ number of nodes during  \texttt{MPI\_COM\_ROW} or \texttt{MPI\_COM\_COL} communications.  Hence, the effective bisection bandwidth, $B_e$, is
\begin{equation}
\label{eq:bisection_bandwidth}
B_e \sim (n')^{2/3} = (n^{1/2})^{2/3} = n^{1/3}.
\end{equation}
In this decomposition, the data per node is $N^3/n$.  During a communication, either row nodes or column nodes are involved.  Hence, the data to be communicated during a \texttt{MPI\_Alltoall} operation is
\begin{equation}
\label{eq:data_per_node}
D = (N^3/n)\times{n^{1/2}} = N^3/n^{1/2}.
\end{equation}
Therefore, the inverse of communication time for each FFT is 
\begin{equation}
T_\mathrm{comm}^{-1} \sim  \frac{B_e}{D} \sim n^{5/6} \approx n^{0.83}.
\label{eq:Blue_Gene_Tcomm_theory}
\end{equation}

We performed our scaling tests on arrays of size $2048^3$, $4096^3$, and $8192^3$ using cores ranging from 256 to 65536.   Each node of Blue Gene/P has 4 cores, hence we performed our simulations on 1, 2, and 4 cores per node, denoted as \texttt{1ppn}, \texttt{2ppn}, and \texttt{4ppn} respectively.     We present all our results in Fig.~\ref{fig:fftk_scaling_bg}, with the subfigures (a,b) exhibiting the inverse of computation and communication timings respectively.  We represent \texttt{1ppn}, \texttt{2ppn}, and \texttt{4ppn}  results using circles, triangles, and squares respectively, and the grid sizes $2048^3$, $4096^3$, and $8192^3$  using increasing sizes of the same symbols.    Fig.~\ref{fig:fftk_scaling_bg} shows that $T^{-1}_\mathrm{comm}$ for $p=256$ and 512 shows better scaling than those for larger number of processors.  This is attributed to the slab decomposition.  Note however that the slab decomposition is possible only when number of processors is smaller than the number of planes of the data ($N$ for $N^3$ grid).   In Fig.~\ref{fig:fftk_scaling_bg_2}(a,b), we exhibit $T^{-1}$ vs. $p$ and $T^{-1}$ vs.~$p/N^3$  to test strong and weak scaling, respectively. 

\setlength{\tabcolsep}{9pt}
\begin{table}[htbp]
\begin{center}
\caption{Comparison of FFTK and P3DFFT on Blue Gene/P for 8192 nodes with \texttt{1ppn} and \texttt{4ppn}. Here $p = $  nodes $\times$ \texttt{ppn}.}
\hspace{20mm}
\begin{tabular}{c c c c }
\hline \hline \\[0.2 pt]

Grid & p & time/step (s) & time/step (s)  \\
        &    &     FFTK         &     P3DFFT \\[2 mm]
\hline \\[0.5 pt]
$4096^3$ &$8192\times 1 $           & $8.18$ & $8.06$ \\[2 mm]
$4096^3$ &$8192 \times 4 $ & $4.14$ & $4.06$ \\[2 mm]
$8192^3$ &$8192\times 1 $          & $71.2$ & $70.0$ \\[2 mm]
$8192^3$ &$8192 \times 4 $ & $45.7$ & $46.2$ \\[2 mm]
\hline
\end{tabular}
\label{tab:comparison_with_p3dfft}
\end{center}
\end{table}

We compute the exponents $\gamma_1$,  $\gamma_2$, and $\gamma$ of Eq.~(\ref{eq:time_sub_division}) using linear regression on the data of Fig.~\ref{fig:fftk_scaling_bg} and Fig.~\ref{fig:fftk_scaling_bg_2}(a).  In Table~\ref{tab:fftk_exponents_bg}, we list the exponents for  \texttt{1ppn}, \texttt{2ppn}, and \texttt{4ppn} and grids sizes of  $2048^3$ and $4096^3$.  As expected, the exponent $\gamma_1 \approx 1$ since the data is equally distributed among all the cores.   The exponent $\gamma_2$ is approximately 0.7 for $2048^3$ for all three cases, but it ranges from 0.8 to 0.9 for $4096^3$.  The increase in $\gamma_2$ with the grid size is probably due to the larger packets communicated for $4096^3$ grids.  The exponent $\gamma_2$ is quite close to the theoretical estimate of $5/6 \approx 0.83$ for $4096^3$ grid [see  Eq.~(\ref{eq:Blue_Gene_Tcomm_theory})].  Our computation also shows the best match for $\gamma_{2}$ with the theoretical estimate is for  \texttt{1ppn}, and it decreases slightly for  for larger \texttt{ppn}.  The variation with \texttt{ppn} is due to {\em cache misses}.

We revisit Fig.~\ref{fig:fftk_scaling_bg_2}(a) that shows a power law scaling,  $T^{-1} \propto p^{\gamma}$,  close to the ideal exponent $\gamma = 5/6$ [see Eq.~(\ref{eq:Blue_Gene_Tcomm_theory})].  This feature is called {\em strong scaling}.  Naturally the larger grids take longer time than the smaller grids.  However, when we increase $p$ and $N^3$ proportionally, all our results collapse into a single curve, as exhibited in $T^{-1}$ vs. $p/N^3$ plot of Fig.~\ref{fig:fftk_scaling_bg_2}(b).  Thus FFTK exhibits both strong and  weak scaling. 

\setlength{\tabcolsep}{16pt}
\begin{table}[htbp]
\begin{center}
\caption{FFTK on Blue Gene/P: Effective FLOP rating in Giga FLOP/s of Blue Gene/P cores for various grid sizes and \texttt{ppn}.   The efficiency $E $ is the ratio of the effective per-core FLOP rating and the peak  FLOP rating of each core (approximately 3.4 Giga FLOP/s).   }
\hspace{20mm}
\begin{tabular}{c c c c }
\hline \hline \\[0.2 pt]

Grid  & \texttt{ppn} &  Giga FLOP/s & $E$ \\ [2 mm]
\hline \\[0.5 pt]
$2048^3$ & 1 &  0.38 & 0.11 \\ 
         & 2 & 0.28 & 0.082 \\ 
         & 4 &  0.17 & 0.050  \\ 
         \hline \\ [2mm]
$4096^3$ & 1 & 0.36 & 0.11 \\ 
         & 2 & 0.25 & 0.073 \\ 
         & 4 &  0.14 & 0.041\\ 
          \hline \\ [2mm]
$8192^3$ & 1 & 0.36 & 0.11 \\ 
         & 2 &    0.26 & 0.076\\
         & 4 &  0.15 & 0.044\\ 
\hline 
\end{tabular}
\label{tab:gflops_bg}
\end{center}
\end{table}

Interestingly $T_\mathrm{comp}$ and $T_\mathrm{comm}$ are comparable on the Blue Gene/P, which is due to the fact that the compute processors are slow, but the interconnect is relatively  fast.  Hence the total time $T$ is impacted by both $T_\mathrm{comp}$ and $T_\mathrm{comm}$.  As a result, the $\gamma$ is reasonably close to unity, thus yielding an approximate linear scaling, at least for the $4096^3$ grid.  We also remark that we have performed FFT for $8192^3$ grid with 8192 and 16384 nodes; it was not possible with lower number of nodes due to memory limitations.  We do not have a reliable scaling exponent for $8192^3$ grid due to lack of data points.

We compare the timings of FFTK with the popular library P3DFFT for $4096^3$ and $8192^3$ grids using 8192 nodes with \texttt{1ppn} and \texttt{4ppn}.  The comparison listed in Table~\ref{tab:comparison_with_p3dfft} indicates that both the libraries are equally efficient.  For comparison we also compute the efficiency of our computation using effective FLOP rating.  A pair of forward and inverse  FFT involves $2 \times 5N^3 \log_2 N^3$ floating point operations~\cite{FFTW05}.  Using this formula we estimate the effective FLOP rating of the supercomputer for various grid sizes and \texttt{ppn}.  The results are listed in Table~\ref{tab:gflops_bg}.   A comparison of the above performance with the average  theoretical rating  of each core (approximate 3.4 Giga FLOP/s)  suggests that the efficiency of the system for a FFT  ranges from approximately 5\% (for 4\texttt{ppn}) to 10\% (for 1\texttt{ppn}) of the peak performance.  The loss of performance is due to large communication time and cache issues during a FFT operation (see Appendix A).    Typically, efficiency of a HPC system  is measured using $T_p/(p T_1)$ where $T_p$ is the time taken to perform operation using $p$ processors.  The data for  large grids, e.g. $1024^3$,  cannot fit in the memory of a single processor, hence we cannot compute $T_1$ and hence $T_p/(p T_1)$.  Therefore we use a more stringent measure.  We measure the efficiency as the ratio of the per-core FLOP rating and the peak rating. We list this efficiency in Table~\ref{tab:gflops_bg}.

In the next subsection we will discuss the scaling of FFTK on Cray XC40.

\subsection{Scaling on Cray XC40}
\label{subsec:FFT_Cray}

Each node of Cray XC40 has 32 compute cores, with each core having an approximate  rating of 36.8 Giga FLOP/s.  Thus each core of Cray XC40 is approximately 10 times faster than that of Blue Gene/P.  Hence, for  given grid size and $p$, $T_{\mathrm{comp}}$ for Cray XC40 is much smaller than that for Blue Gene/P.

 The Cray XC40 employs dragonfly topology which consists of hierarchy of structures that yields bandwidth proportional to the number of interacting nodes.  Hence the bandwidth is
\begin{equation}
B_e \sim n', 
\end{equation} 
where $n'$ is the number of interacting nodes.  For the pencil decomposition, the total number of nodes $n$ is divided as  $n = n_\mathrm{row} \times n_\mathrm{col}$.   Hence $n' = n/n_\mathrm{row}$ for \texttt{MPI\_COMM\_COL} communicator and  $n' = n/n_\mathrm{col}$ for \texttt{MPI\_COMM\_ROW} communicator.   Note that the data to be communicated during \texttt{MPI\_Alltoall} operation is $(N^3/n)n'$.  Hence, the inverse of the communication time is
\begin{equation}
T_\mathrm{comm}^{-1} \sim  \frac{B_e}{D} \approx  \frac{n'}{(N^3/n) n'} \sim n,
\label{eq:T_comm_XC40}
\end{equation}
implying a linear scaling.  

Comparison of $T_\mathrm{comm}$ for Blue Gene/P and Cray XC40 reveal that the bandwidth is larger for Cray XC40 than Blue Gene/P.  Hence, for a given set of $N,n$, and $D$ (the data to be communicated), the time for communication is smaller for Cray XC40 than Blue Gene/P (see Eqs.~(\ref{eq:Blue_Gene_Tcomm_theory}, \ref{eq:T_comm_XC40})).  Thus, the data communication in XC40 is more efficient than that in Blue Gene/P.  However, we will show later that the gain in the speed of the  interconnect is in commensurate with the increase in the computational power of  the processor.  Hence, for the FFT computation, the overall efficiency of Cray XC40 is lower than that for Blue Gene/P. We also remark that Hadri {\em et al.}~\cite{Bilel:shaeen2} showed that the maximum global bandwidth between the 18 groups of two cabinets is approximately 57\% of the peak performance, hence we expect suboptimal performance for communications for FFT due to  \texttt{MPI\_Alltoall} data exchange.

\begin{figure}[htbp]
\begin{center}
\includegraphics[scale = 0.8]{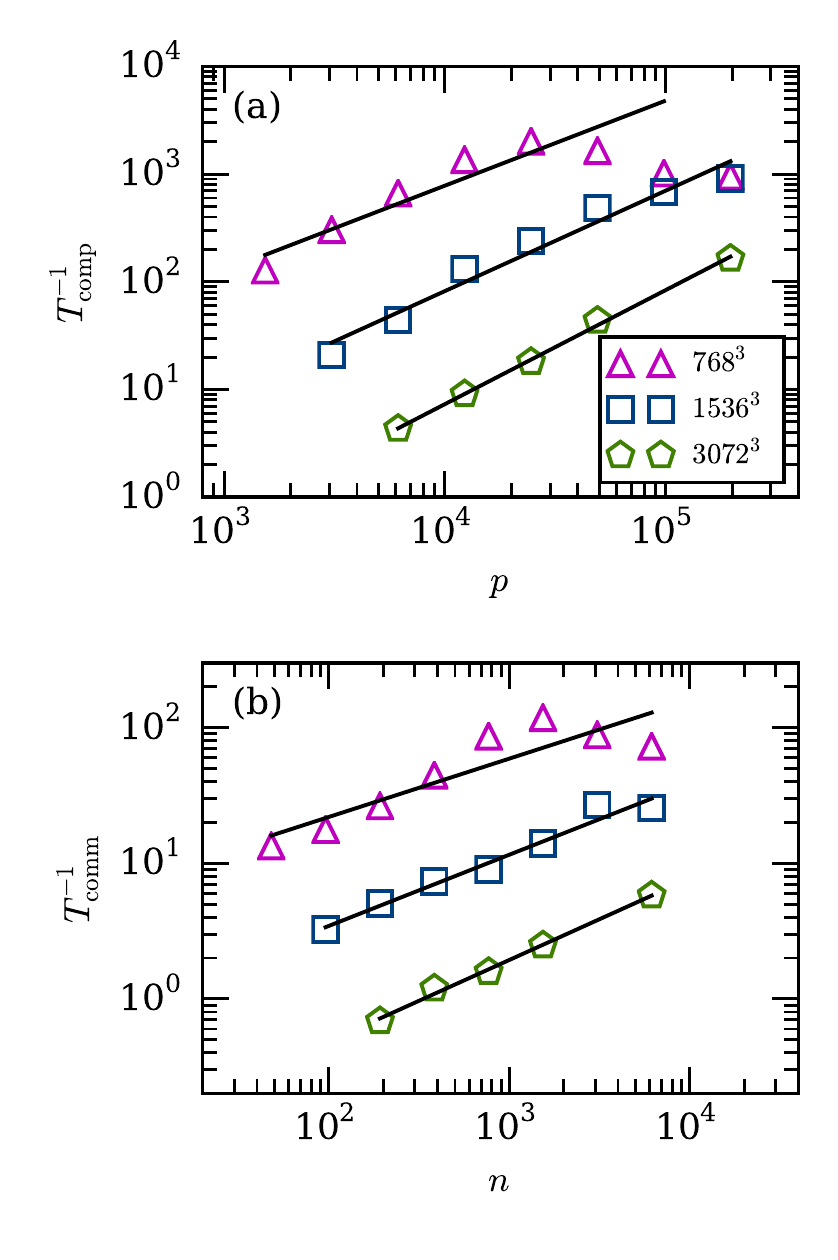}
\end{center}
\caption{Scalings of FFTK on Cray XC40 for the \texttt{FFF} basis: (a) Plots of  $T^{-1}_\mathrm{comp}$ vs.~$p$  (number of cores) for $768^3$, $1536^3$, and $3072^3$  grids.  (b) Plots of $T^{-1}_\mathrm{comm}$ vs.~$n$ (number of nodes) using the above convention.}
\setlength{\abovecaptionskip}{0pt}
\label{fig:fftk_scaling_cray}
\end{figure}

\begin{figure}[htbp]
\begin{center}
\includegraphics[scale = 0.8]{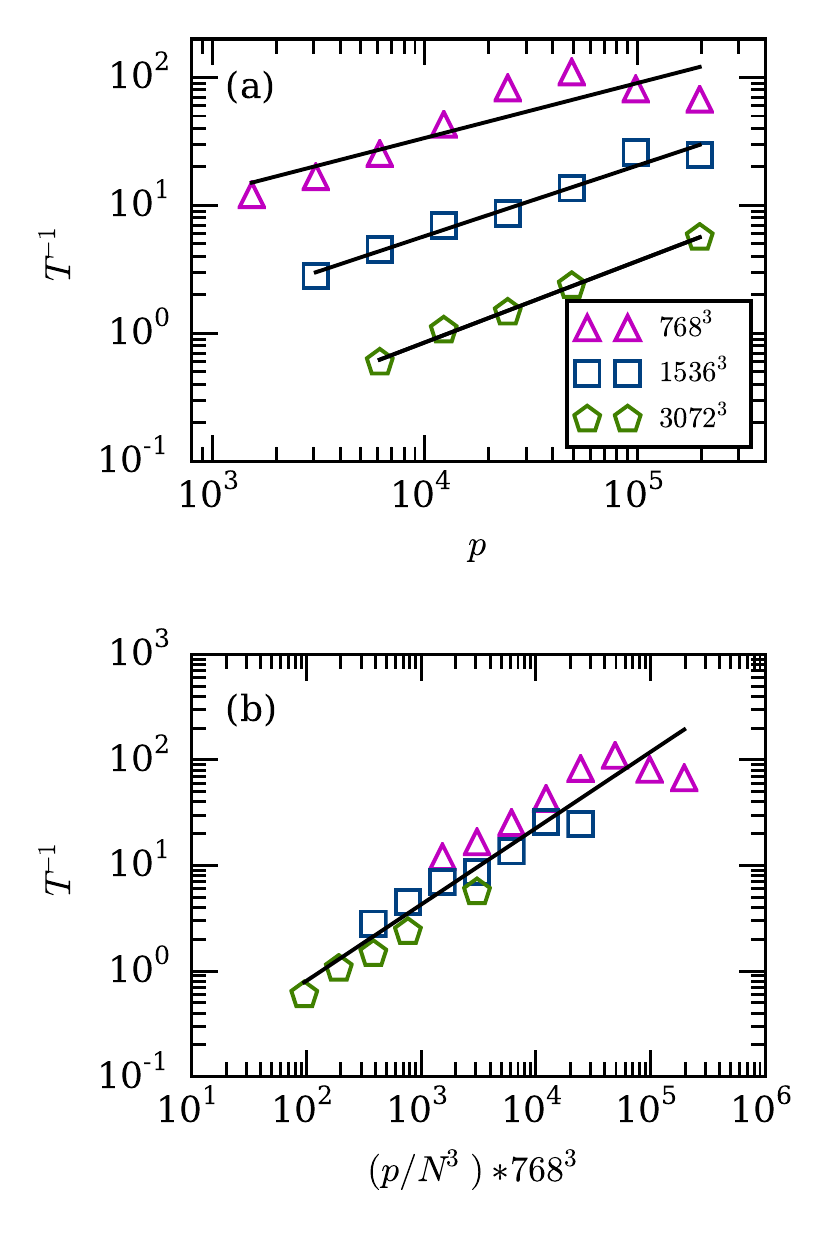}
\end{center}
\setlength{\abovecaptionskip}{0pt}
\caption{Scalings of FFTK on Cray XC40 for the \texttt{FFF} basis: (a) plots of  $T^{-1}$ vs.~$p$  for $768^3$, $1536^3$, and $3072^3$ grids.  (b) plots of $T^{-1}$ vs.~$p/N^3$ exhibits weak scaling with an exponent of $\gamma=0.72\pm0.03$.}
\label{fig:fftk_scaling_cray_2}
\end{figure}

\setlength{\tabcolsep}{10pt}
\begin{table}[htbp]
\begin{center}
\caption{FFTK scaling on Cray XC40 for the \texttt{FFF} and  \texttt{SFF} basis:   The exponents $\gamma_1$ for the computation time ($T_\mathrm{comp}$), $\gamma_2$ for the communication time  ($T_\mathrm{comm}$), and $\gamma$ for the total time  ($T$) [refer to Eq.~(\ref{eq:time_sub_division}) for definition].  Maximum cores  used: 196608.}
\hspace{20mm}
\begin{tabular}{c c c c }
 \hline \\[0.2 pt]
\multicolumn{4}{c}{\texttt{FFF}} \\ [0.5mm]
\hline \hline \\[0.2 pt]
$\mathrm{Grid}$ &  $\gamma_1$ &  $\gamma_2$ &  $\gamma$ \\[2 mm]
\hline \\[0.2 pt]
$768^3$ & $0.79 \pm 0.14$ & $0.43 \pm 0.09$  & $0.43 \pm 0.09$ \\
 \\[0.2 pt]
$1536^3$ & $0.93 \pm 0.08$ & $0.52 \pm 0.04$  & $0.55 \pm 0.04$ \\
 \\[0.2 pt]
$3072^3$        & $1.08 \pm 0.03$ & $0.60 \pm 0.02$  & $0.64 \pm 0.02$ \\
 \hline \\[0.2 pt]
\multicolumn{4}{c}{ \texttt{SFF}} \\ [0.5mm]
\hline \hline \\[0.2 pt]
$\mathrm{Grid}$ &  $\gamma_1$ &  $\gamma_2$ &  $\gamma$ \\[2 mm]
\hline \\[0.2 pt]
$768^3$ & $0.82 \pm 0.13$ & $0.44 \pm 0.03$  & $0.46 \pm 0.04$ \\
\\[0.2 pt]
$1536^3$ & $0.97 \pm 0.07$ & $0.63 \pm 0.02$  & $0.66 \pm 0.01$ \\
 \\[0.2 pt]
$3072^3$        & $0.99 \pm 0.04$ & $0.70 \pm 0.05$  & $0.73 \pm 0.05$ \\
\hline
\end{tabular}
\label{tab:fftk_exponents_cray}
\end{center}
\end{table}

We performed FFTK scaling on grids of sizes $768^3$ to $6144^3$ using cores ranging from 1536 to 196608 ($3\times 2^{16}$).  Each node contains 32 cores,  which implies a somewhat large number of \texttt{ppn} combinations.  Hence we choose to use all the cores in a given node for maximum utilization.  Both the grid sizes and number of cores are of the form $3\times 2^n$ since the maximum number of cores in Cray XC40, $3\times 2^{16}$, is divisible by 3.  

We perform scaling analyses for the \texttt{FFF} and  \texttt{SFF} basis.   However we present our results on $T_\mathrm{comp}$, $T_\mathrm{comm}$, and $T$ in Figs.~\ref{fig:fftk_scaling_cray} and \ref{fig:fftk_scaling_cray_2} for the \texttt{FFF} basis only since  \texttt{SFF} basis has similar behavior.  We observe that the total computation time for the process is an order of magnitude smaller than the total communication time.  Hence the efficiency of FFT is dominated by the \texttt{MPI\_Alltoall} communication of the FFT.  The figures show that $T_\mathrm{comp}^{-1} \sim p^{\gamma_1} $, $T_\mathrm{comm}^{-1} \sim n^{\gamma_2}$, and $T^{-1} \sim p^{\gamma} $, with  minor deviations from the power law arising for $768^3$ grid with large $p$'s ($p \ge 98000$).  Thus the data exhibits a strong scaling nearly up to 196608 cores.    We also observe that all the data nearly collapse to a single curve when we plot $T^{-1}$ vs.~$p/N^3$, hence FFTK exhibits both weak and strong scaling nearly up to 196608 cores.

\setlength{\tabcolsep}{17pt}
\begin{table}[htbp]
\begin{center}
\caption{FFTK on Cray XC40: Effective FLOP rating in Giga FLOP/s of Cray XC40 cores for various grid sizes and \texttt{ppn}. The efficiency $E$  is the ratio of the effective per-core FLOP rating and the peak FLOP rating of each core (approximately 36 Giga FLOP/s).  }
\hspace{20mm}
\begin{tabular}{c c c c }
\hline \hline \\[0.2 pt]
Grid Size & $768^3$ & $1536^3$ & $3072^3$  \\[2 mm]
  GFlop/s &  0.45   &   0.53   &   0.64    \\[2 mm]
$E$ & 0.013    &   0.015   &   0.018    \\[2mm]
\hline
\end{tabular}
\label{tab:gflops-cray}
\end{center}
\end{table}

The exponents for various grids for the  \texttt{FFF} and  \texttt{SFF} basis are listed in Table~\ref{tab:fftk_exponents_cray}.  We observe that $\gamma_1 \approx 1$, except for $768^3$, thus yielding a linear scaling for the computation time.   The exponent $\gamma_2$ of the communication time ranges from 0.43 to 0.70 as we increase the grid size from $768^3$ to $3072^3$, which may be due to increased efficiency of the network for larger data size.  As described above, the total time is dominated by the communication time, hence $\gamma \approx \gamma_2$.    Also, the  \texttt{SFF} basis appears to be slightly more efficient than the  \texttt{FFF} basis.

We compute the effective FLOP rating of the supercomputer by dividing the total number of floating operations for a pair of FFT ($2\times 5 N^3 \log_2 N^3$) with the total time taken. This operation yields the average rating of each core of Cray XC40 as 0.45 to 0.64 Giga FLOP/s that translates to 1.3\% to 1.8\% of the peak performance (36 Giga FLOP/s) (see Table~\ref{tab:gflops-cray}). 

 It is important to contrast the efficiencies of the two HPC supercomputers discussed in this paper.  The efficiency of Blue Gene/P  at approximately $4\%$ (for 4\texttt{ppn}) is higher than that of Cray XC40 ($\sim1.5\%$).  A node of Cray XC40 comprises of 32 cores, each with a peak rating of 36 Giga FLOP/s rating. Hence maximum  compute power per node is $1177.6$ Giga FLOP/s.  On the other hand, a Blue Gene/P node contains 4 cores with a peak rating of $4\times 3.4 = 13.6$ Giga FLOP/s.  Thus, each node of Cray XC40  has approximately 100 times more computational power.  However, the interconnect of Cray XC40 is not faster than that of Blue Gene/P in the same ratio.
 
 Note that the dragonfly topology of Cray XC40 appears to be more efficient than the torus topology of Blue Gene/P (see Eqs.~(\ref{eq:Blue_Gene_Tcomm_theory}),~(\ref{eq:T_comm_XC40})), yet the ratio of the speedup is much less than 100.    Therefore, for data communication the processors of Cray have to idle longer than those of Blue Gene/P.  For the Blue Gene/P, the relatively slower processors do not have to idle as long. This is especially critical for FFT which is communication intensive.  Thus, the faster processor and relatively slower interconnect of Cray XC40 result in an overall lower efficiency compared to Blue Gene/P.  This is essentially the reason for the lower efficiency of Cray XC40 at 1.5\% compared with 4\% of the Blue Gene/P.  In this sense,  the efficiency of hardware depends on the application;  for FFT computation, a faster switch is more important than a faster processor.

\subsection{{Comparison between FFTK and P3DFFT libraries}}
{In this section we describe key features and performance of FFTK library.  Another library P3DFFT has similar features.  Therefore, it is important to compare the features and performances of the two FFT libraries, which are briefly described below:}
\begin{enumerate}

{

\item P3DFFT library has functions to perform Fourier transforms along the three directions. It also has functions to perform  Sine transform, Cosine transform or Chebyshev transform only along one direction, and  Fourier transforms along the rest two directions~\cite{p3dfft:user}. Thus P3DFFT can be used for mixed basis functions like \texttt{SFF} and \texttt{ChFF} (see Sec.~\ref{sec:Num_Sch} for definitions).  In contrast, FFTK can perform, Fourier, Sine, or Cosine transforms along any of the three directions.  Thus it can be used for solving problems in  \texttt{FFF},  \texttt{SFF},  \texttt{SSF}, and \texttt{SSS} basis.   Hence, FFTK has more features than P3DFFT.  \texttt{SSS} is very import basis function for physical application, like, reversal studies of Rayleigh--B\'{e}nard convection~\cite{Verma:PF2015Reversal}. The FFTK library does not yet support \texttt{ChFF} basis function, but it is under developement.

\item In the present paper, we report scaling of FFTK on Cray XC40 for number of processors up to 196608.  The P3DFFT library has been scaled up to 65536 cores~\cite{Pekurovsky:SC12} on Cray XT5 machine.  To best of our knowledge, no other group has performed detailed scaling studies on FFT on processors more than that for FFTK.  Many researchers have performed spectral simulations of fluid flows on a large number of cores, for example, Yeung~{\em et al.}~\cite{Yeung:PNAS2015} used $262144$ cores for their $8192^3$ simulation, but they did not report scaling results of FFT. Note that, scaling study of FFT is important for the optimised performance of spectral codes. 

\item As we show in this paper, the performance of FFT depends critically on the features of processors as well as that of interconnect.  Here we perform comparative study of FFTK on BlueGene/P and Cray XC40.  One of our findings is that for FFT computations, the efficiency  of Cray XC40 is lower than BlueGene/P even though Cray XC40 is more modern HPC system than Blue Gene/P.  This is because the efficiency of interconnects has not grown in commensurate with that of processors.  We are not aware of similar extensive comparisons and analysis of scaling results for FFT. 

\item As shown in Table~\ref{tab:comparison_with_p3dfft}, on BlueGene/P, FFTK and P3DFFT are equally efficient.}
\end{enumerate}

{ After extensive discussion on FFTs, in the next section, we will present the scaling results of the fluid spectral solver.}

\section{Scaling of Fluid spectral solver}
\label{sec:result_fluid}

We perform high-resolution fluid simulation using spectral method  on Blue Gene/P and Cray XC40, i.e., we solve Eq.~(\ref{eq:NS},\ref{eq:continuity1}) on these systems.  We assume the flow to be incompressible, and use periodic boundary condition for which  \texttt{FFF} basis function is appropriate.  

The  fluid simulation requires 15  arrays each of size $N^3$ to store three components of the velocity field  in real and Fourier spaces, the force field, the nonlinear term ${\bf u} \cdot \nabla {\bf u}$, and three temporary arrays~\cite{Verma:Pramana2013}.  We employ the fourth-order Runge Kutta scheme for time stepping, and dealias the nonlinear terms using the 2/3-rd rule.  Each time step requires 36 FFT operations.   We refer the reader to Boyd~\cite{Boyd:Book}, Canuto {\em et al.}~\cite{Canuto:SpectralFluid_book}, and Verma {\em et al.}~\cite{Verma:Pramana2013} for further details and validation tests of the fluid solver.    We run our simulations for 10 to 100 time steps depending on the grid size.  The time reported in the present section is an average over these many time steps.

\setlength{\tabcolsep}{9pt}
\begin{table}[htbp]
\begin{center}
\caption{Scaling exponent of the total time for the fluid solver on  Blue Gene/P and Cray XC40 for various grids (definition: $T \sim p^{{-\gamma}}$).}
\hspace{20mm}
\begin{tabular}{  c c   c c}
\hline \hline \\[0.3 pt]
                    \multicolumn{2}{c}{Blue Gene/P} & \multicolumn{2}{c}{Cray XC40} \\
                   \hline \\[0.2 pt]
Grid size & $\gamma$ & Grid size & $\gamma$ \\ [2 mm]
\hline \\[0.5 pt]
$2048^3$    & $0.95 \pm 0.05$  & $768^3$  & $0.28  \pm 0.15$  \\[2 mm]
$4096^3$    & $0.8 \pm 0.1$    & $1536^3$ & $0.44 \pm 0.06$ \\[2 mm]
                        -        &      -          & $3072^3$ & $0.68 \pm 0.02$ \\[2 mm]
\hline
\end{tabular}
\label{tab:scaling_exponents_fluid}
\end{center}
\end{table}

The most expensive part of a pseudospectral simulation is the FFT that consumes approximately 70\% to 80\% of the total time.  In addition, a flow solver has many functions including I/O, hence it is tedious to find patterns for the computation and communication components separately.  Therefore,  in the following discussion we report the scaling of the total time for the flow solvers.

\subsection{Blue Gene/P}
\begin{figure}[htbp]
\begin{center}
\includegraphics[scale = 0.8]{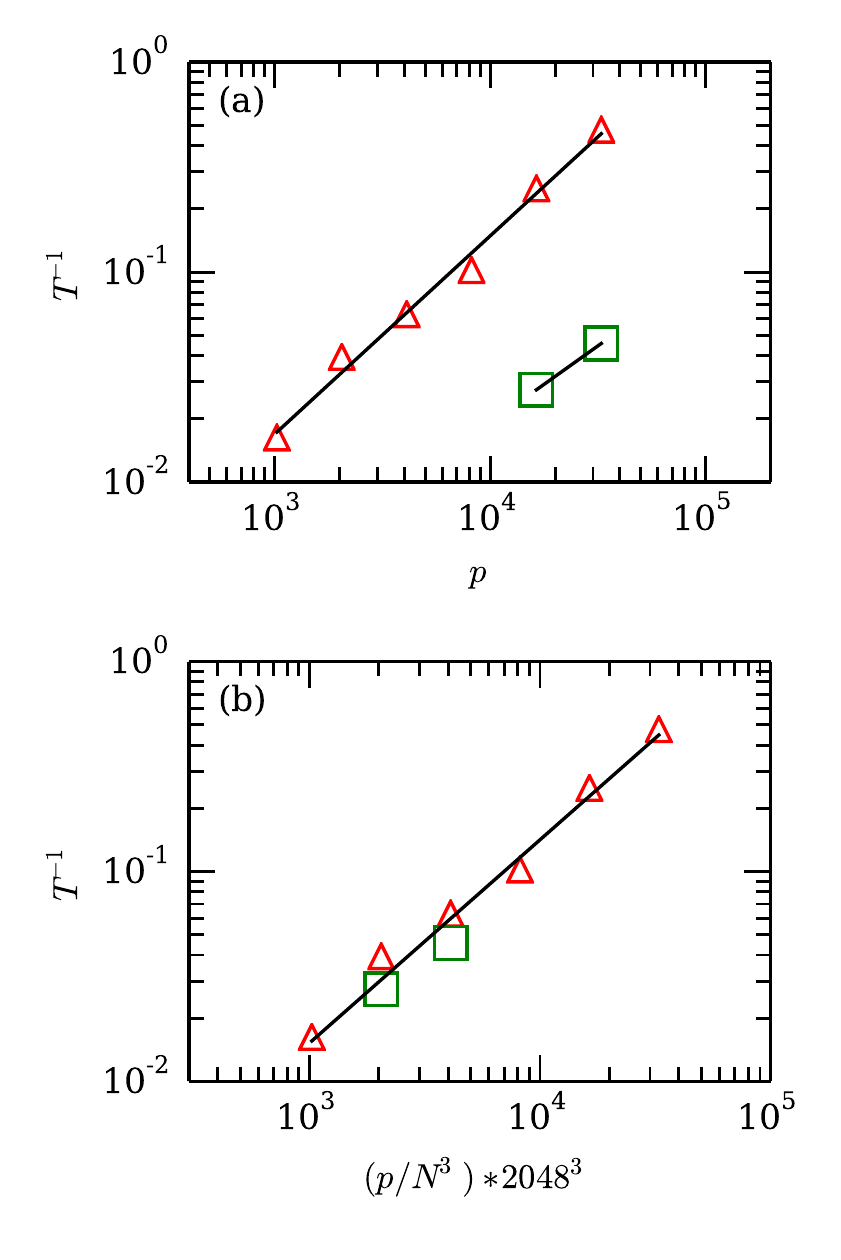}
\end{center}
\setlength{\abovecaptionskip}{0pt}
\caption{Scaling of the fluid spectral solver on Blue Gene/P: (a) Plot of  $T^{-1}$ vs.~$p$ for $2048^3$ (red triangle) and $4096^3$ (green square) grids exhibits strong scaling. (b)  Plot of $T^{-1}$ vs.~$p/N^3$ exhibits weak scaling with an exponent of $\gamma=0.97\pm0.06$.}
\label{fig:fluid_scaling_bg}
\end{figure}

We performed the fluid simulation on $2048^3$ and $4096^3$ grids using cores ranging from 1024 to 65536.   In Fig.~\ref{fig:fluid_scaling_bg}(a) we plot the inverse of the total time per iteration vs.~$p$.  We observe that $T^{-1} \sim p^\gamma$.  The exponents $\gamma$  listed in Table~\ref{tab:scaling_exponents_fluid} shows that $\gamma = 0.95\pm0.05$ and $\gamma = 0.8\pm0.1$ for grid sizes of $2048^3$ and $4096^3$ respectively.  This demonstrates  the fluid solver exhibits a strong scaling.

The above data nearly collapses into a single curve in the  plot of $T^{-1}$ vs. $p/N^3$, as shown in Fig.~\ref{fig:fluid_scaling_bg}(b).  Hence we conclude that our fluid solver also exhibits weak scaling.  The exponent of the weak scaling is $\gamma = 0.97 \pm 0.06$. 

\subsection{Cray XC40}

We performed fluid simulations on grid sizes of $768^3$, $1536^3$ and $3072^3$ grids using  cores ranging from 1536 to 196608. We employ periodic boundary conditions along all the walls. As done for Blue Gene/P supercomputer, we compute the total time taken for each iteration of the solver.  

\begin{figure}[htbp]
\begin{center}
\includegraphics[scale = 0.8]{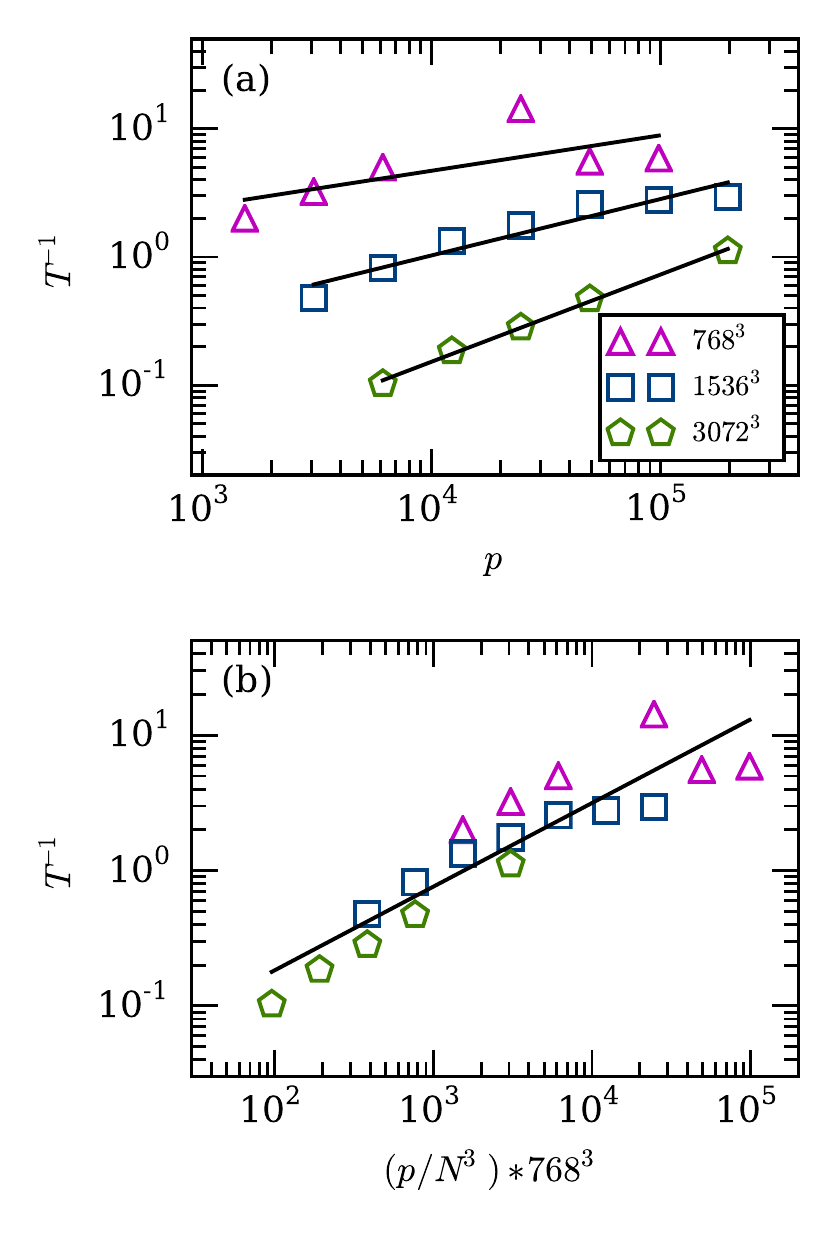}
\end{center}
\setlength{\abovecaptionskip}{0pt}
\caption{Scaling of the fluid spectral solver on Cray XC40: (a) Plot of $T^{-1}$ vs.~$p$ for $768^3$, $1536^3$, and $3072^3$  grids exhibits strong scaling. (b)  Plot of $T^{-1}$ vs.~$p/N^3$ exhibits weak scaling with an exponent $\gamma=0.62\pm0.07$.}
\label{fig:fluid_scaling_cray}
\end{figure}

Fig.~\ref{fig:fluid_scaling_cray}(a)  shows that the plots of $T^{-1} \propto p^\gamma$, except for $768^3$ grid with large $p$'s ($p \ge 98000$).  Thus fluid solver exhibits a strong scaling, except for $768^3$ grid that scales up to $p \lessapprox 98000$.  We observe that $\gamma$ for the $768^3$, $1536^3$ and $3072^3$ grids are approximately  0.28, 0.44 and 0.68 respectively.  The three curves for the three different grids collapse into a single curve when the $X$-axis is chosen as $p/N^3$   (except for $p \ge 98000$).  This result shows a common scaling when the number of cores and data sizes are increased by an equal factor.  Thus our fluid solver  exhibits both weak and strong scaling  nearly up to 196608 cores of the Cray XC40.

 We performed a fluid simulation for a longer time on a  $512^{3}$ grid in a  periodic box of size $(2\pi)^3$.  The Reynolds number of the flow at the steady state is approximately $1100$.  Fig.~\ref{fig:fluid_fig} exhibits an isosurface of the contours of constant  magnitudes of the vorticity under steady state.  In the figure we observe intense localised vorticity, as reported in literature.

\section{Scaling of RBC spectral solver}
\label{sec:result_rbc}

We performed high-resolution simulations of Rayleigh-B\'{e}nard convection (RBC) by solving Eqs.~(\ref{eq:u_Usmall_Plarge},~\ref{eq:T_Usmall_Plarge}).  The fluid is assumed to be contained in a box of unit dimension.  Presently, we report the scaling results for  \texttt{FFF} (periodic boundary condition) and  \texttt{SFF} (free-slip boundary condition) basis functions.   Note that  in  \texttt{SFF} basis, $u_z=\partial_z u_x = \partial_z u_y = 0$ at the top and bottom plates, and periodic along the side walls.  The temperatures at the top and bottom plates are assumed to be constant (conducting walls), while at the side walls, the temperature is assumed to be  periodic.    For the energy spectrum and flux computations, we employ a free-slip boundary condition.

A RBC simulation requires $18$ arrays of size $N^3$.  We time step the solver using the fourth-order Runge Kutta scheme that requires 52 FFT  per time step.   For further details and validation tests of the RBC solver, we refer  the reader to Verma {\em et al.}~\cite{Verma:Pramana2013}.  For scaling tests we run our simulations for 10 to 100 time steps.  The time reported in this section is an average over the relevant time steps.  The results of our simulations on the Blue Gene/P and Cray XC40 supercomputers are as follows:

 \begin{figure}[htbp]
\begin{center}
\includegraphics[scale = 0.8]{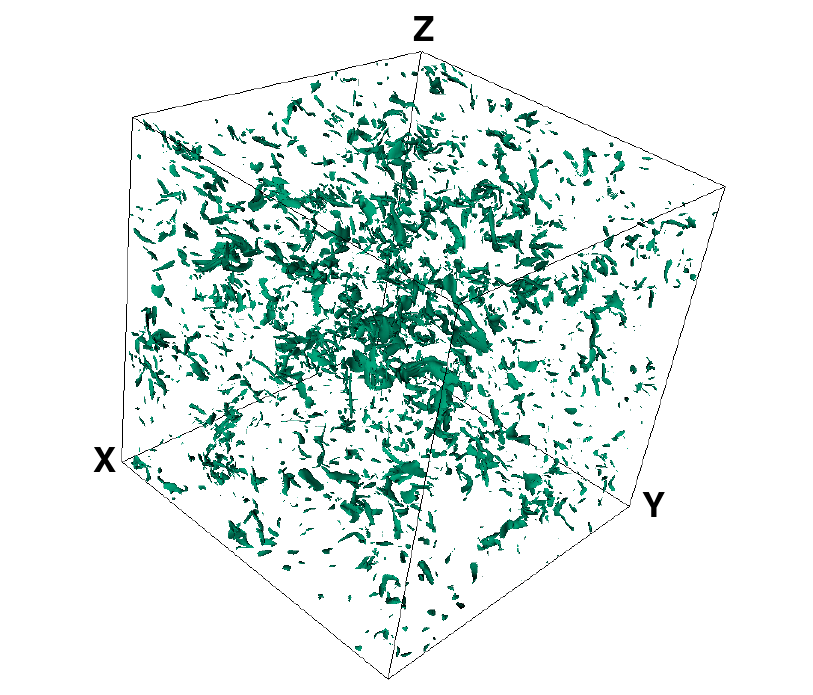}
\end{center}
\setlength{\abovecaptionskip}{0pt}
\caption{Isosurface of the contours of constant vorticity $|\nabla \times {\bf u}|$ ($30 \%$ of the maximum value).  The figure indicates regions of strong vorticity in the flow. }
\label{fig:fluid_fig}
\end{figure}

\subsection{Blue Gene/P}
\begin{figure}[htbp]
\begin{center}
\includegraphics[scale = 0.78]{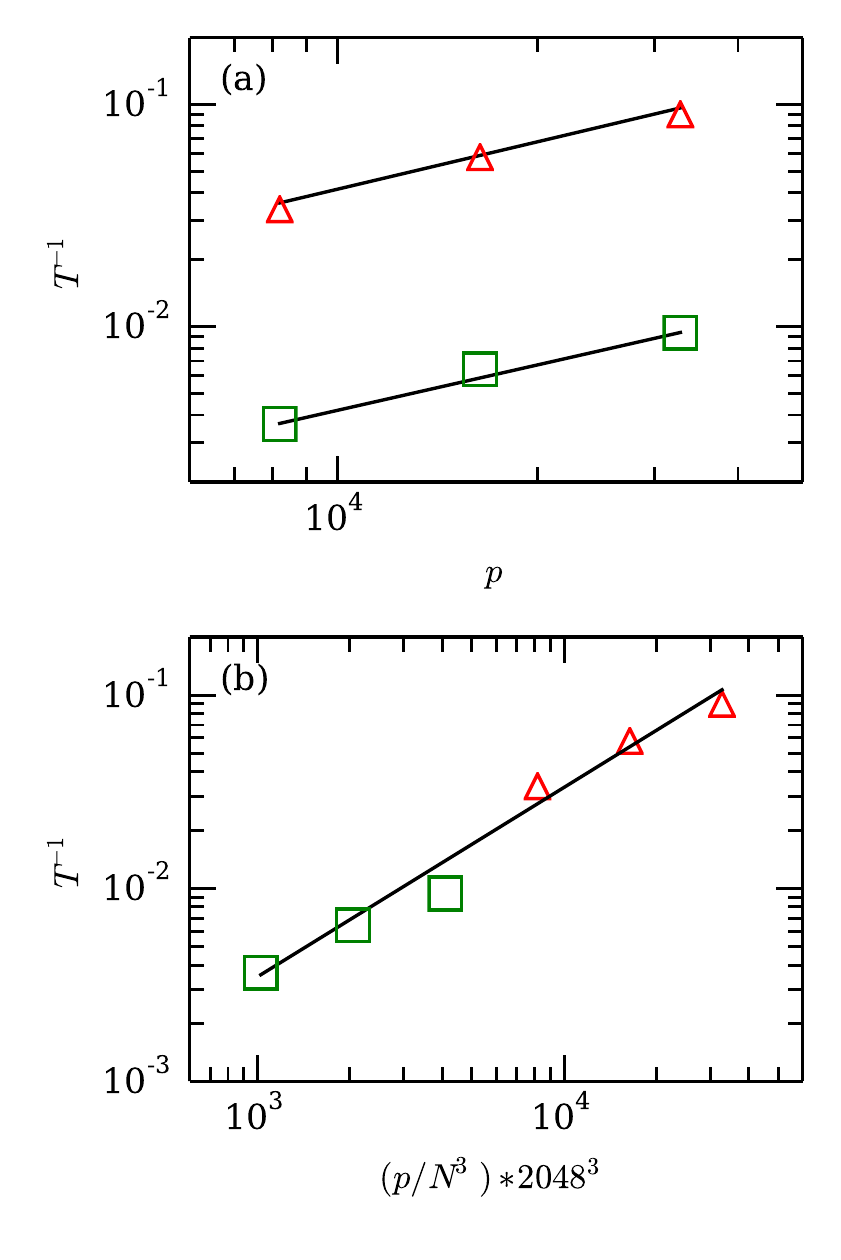}
\end{center}
\setlength{\abovecaptionskip}{0pt}
\caption{Scaling of the RBC solver on Blue Gene/P for the  \texttt{FFF} basis:  (a) Plot of  $T^{-1}$ vs.~$p$ for $2048^3$ (red triangle) and $4096^3$ (green square) grids exhibits strong scaling. (b)  Plot of $T^{-1}$ vs.~$p/N^3$ exhibits weak scaling with an exponent of $\gamma=0.68\pm0.08$.}
\label{fig:RBC_scaling_bg}
\end{figure}

We performed RBC simulations on $2048^3$ and $4096^3$ grids using cores ranging from $8192$ to $65536$. In Fig.~\ref{fig:RBC_scaling_bg}(a,b)  we plot $T^{-1}$ vs.~$p$  and $T^{-1}$ vs.~$p/N^{3}$ respectively.  Here $T$ is the time taken per step, and $p$ is the number of cores.  We observe that $T^{-1} \sim p^\gamma$ with the exponent $\gamma = 0.71$ and $0.68$ for the $2048^3$ and $4096^3$ grids  respectively (see Table~\ref{tab:scaling_exponents_RBC}).  Thus the RBC solver indicating a strong scaling.   As exhibited in Fig.~\ref{fig:RBC_scaling_bg}(b), the data nearly collapses into a single curve when we plot $T^{-1}$ vs.~$p/N^3$, hence exhibiting a weak scaling as well.

\setlength{\tabcolsep}{6pt}
\begin{table}[htbp]
\begin{center}
\caption{Scaling exponents of the total time for the RBC solver on Blue Gene/P and Cray XC40 for various grids for the  \texttt{FFF} and  \texttt{SFF} basis functions  (definition: $T \sim p^{{-\gamma}}$).}
\begin{tabular}{c c c c}
\\[0.5mm]
\hline \\[0.5mm]
                   \multicolumn{2}{c}{Blue Gene/P} & \multicolumn{2}{c}{Cray XC40} \\[0.5mm]
\hline \\[0.2pt]
\multicolumn{4}{c}{ \texttt{FFF}} \\[0.5mm]       
                   \hline \hline \\[0.2 pt]    
             Grid Size & $\gamma$ & Grid Size & $\gamma$ \\ [2 mm]
\hline \\[0.5 pt]    
				 $2048^3$    & $0.71 \pm 0.04$  & $768^3$  & $0.49  \pm 0.14$  \\[2 mm]
             $4096^3$    & $0.68 \pm 0.08$  & $1536^3$ & $0.64 \pm 0.04$ \\[2 mm]
                        -       &        -         & $3072^3$ & $0.74 \pm 0.03$ \\[2 mm]
\hline \\
\multicolumn{4}{c}{ \texttt{SFF}} \\ [0.5mm]       
                   \hline\hline \\[0.2 pt]    
                                Grid Size & $\gamma$ & Grid Size & $\gamma$ \\ [2 mm]
\hline \\[0.5 pt]   
				 -   & -  & $768^3$  & $0.62  \pm 0.06$  \\[2 mm]
             -   &-  & $1536^3$ & $0.74 \pm 0.09$ \\[2 mm]
                        -       &        -         & $3072^3$ & $0.80 \pm 0.05$ \\[2 mm]
\hline
\end{tabular}
\label{tab:scaling_exponents_RBC}
\end{center}
\end{table}

\subsection{Cray XC40}

\begin{figure}[htbp]
\begin{center}
\includegraphics[scale = 0.78]{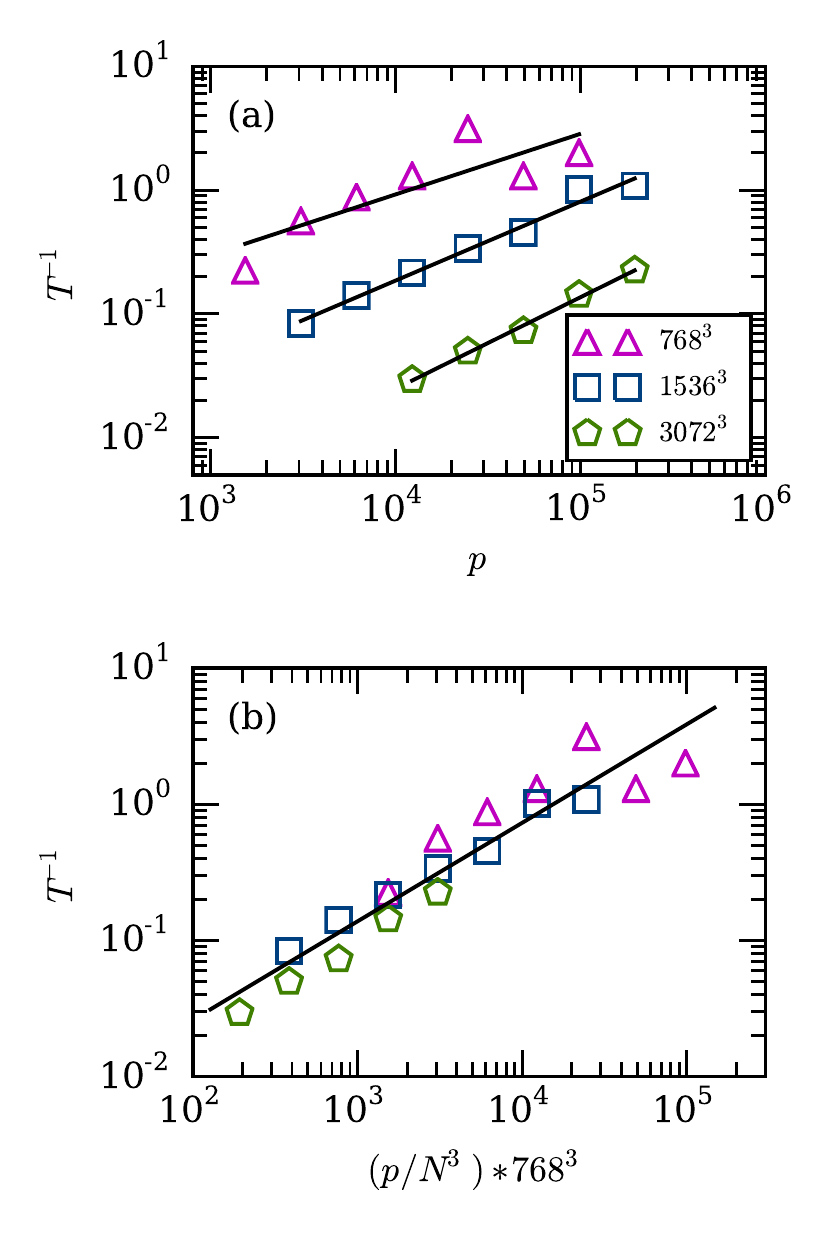}
\end{center}
\setlength{\abovecaptionskip}{0pt}
\caption{Scaling of the RBC spectral solver for the \texttt{FFF} basis on Cray XC40: (a) Plot of  $T^{-1}$ vs.~$p$ for $768^3$, $1536^3$, and $3072^3$ grids exhibits strong scaling. (b)  Plot of $T^{-1}$ vs.~$p/N^3$ exhibits weak scaling with an exponent of $\gamma=0.72\pm0.06$.}
\label{fig:RBC_scaling_cray}
\end{figure}

\begin{figure}[htbp]
\begin{center}
\includegraphics[scale = 0.78]{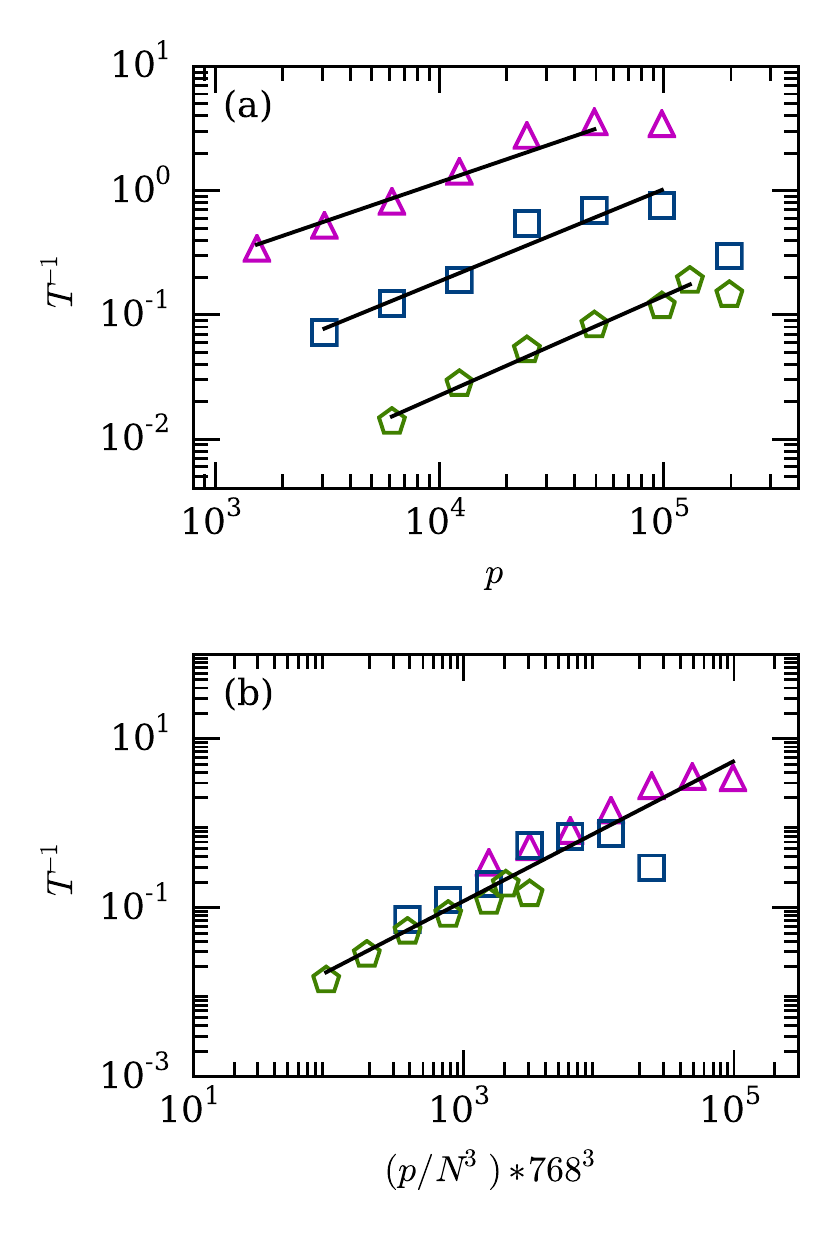}
\end{center}
\setlength{\abovecaptionskip}{0pt}
\caption{ Scaling of the RBC spectral solver for the \texttt{SFF} basis on Cray XC40: (a) Plot of  $T^{-1}$ vs.~$p$ for $768^3$, $1536^3$, and $3072^3$ grids exhibits strong scaling. (b)  Plot of $T^{-1}$ vs.~$p/N^3$ exhibits weak scaling with an exponent of $\gamma=0.83\pm0.03$.}
\label{fig:RBC_scaling_cray_SFF}
\end{figure}

 We simulated RBC on  $768^3$, $1536^3$ and $3072^3$ using  cores ranging from 1536 to 168608 for  \texttt{FFF} and  \texttt{SFF} basis. For 3300 iterations of RBC simulation on $2048^3$ grid, the total simulation is $1.7 \times 10^5$ seconds, thus the time per iteration of RBC on $2048^3$ grid is approximately 51.5 seconds.  For the FFF basis, the inverse of the total time plotted in Fig.~\ref{fig:RBC_scaling_cray}(a) scales as $T^{-1} \sim p^\gamma$ with $\gamma = 0.49$, 0.64 and 0.74 for  $768^3$ (except for $p =196608 $), $1536^3$ and $3072^3$ grids respectively (also see Table~\ref{tab:scaling_exponents_RBC}).  The plot indicates a strong scaling for the RBC solver.   In Fig.~\ref{fig:RBC_scaling_cray}(b) we plot $T^{-1}$ vs. $p/N^3$; here  the data collapses into a single curve (see Fig.~\ref{fig:RBC_scaling_cray}) thus indicating a weak scaling.  In Fig.~\ref{fig:RBC_scaling_cray_SFF} we plot $T^{-1}$ vs.~$p$ and $T^{-1}$ vs.~$p/N^3$ for the  \texttt{SFF} basis.  The exponents listed in Table~\ref{tab:scaling_exponents_RBC} show that the  \texttt{FFF} and  \texttt{SFF} scale in a similar manner, with the \texttt{SFF} basis scaling slightly better than the \texttt{FFF} basis. The plots and the scaling exponents demonstrate that our RBC solver exhibits both strong and weak scaling  up to nearly 196608 cores.

\begin{figure}[htbp]
\begin{center}
\includegraphics[scale = 0.78]{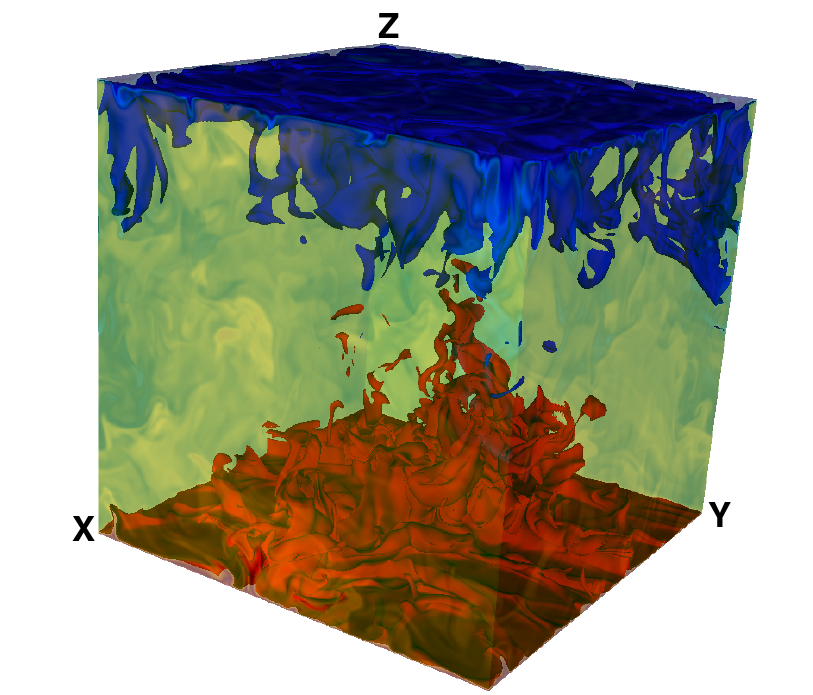}
\end{center}
\setlength{\abovecaptionskip}{0pt}
\caption{Isosurface of the contours of constant temperature.  The structures with red and blue colours indicate respectively the hot and cold plumes of the flow.  }
\label{fig:RBC_fig}
\end{figure}

\begin{figure}[htbp]
\begin{center}
\includegraphics[scale = 0.78]{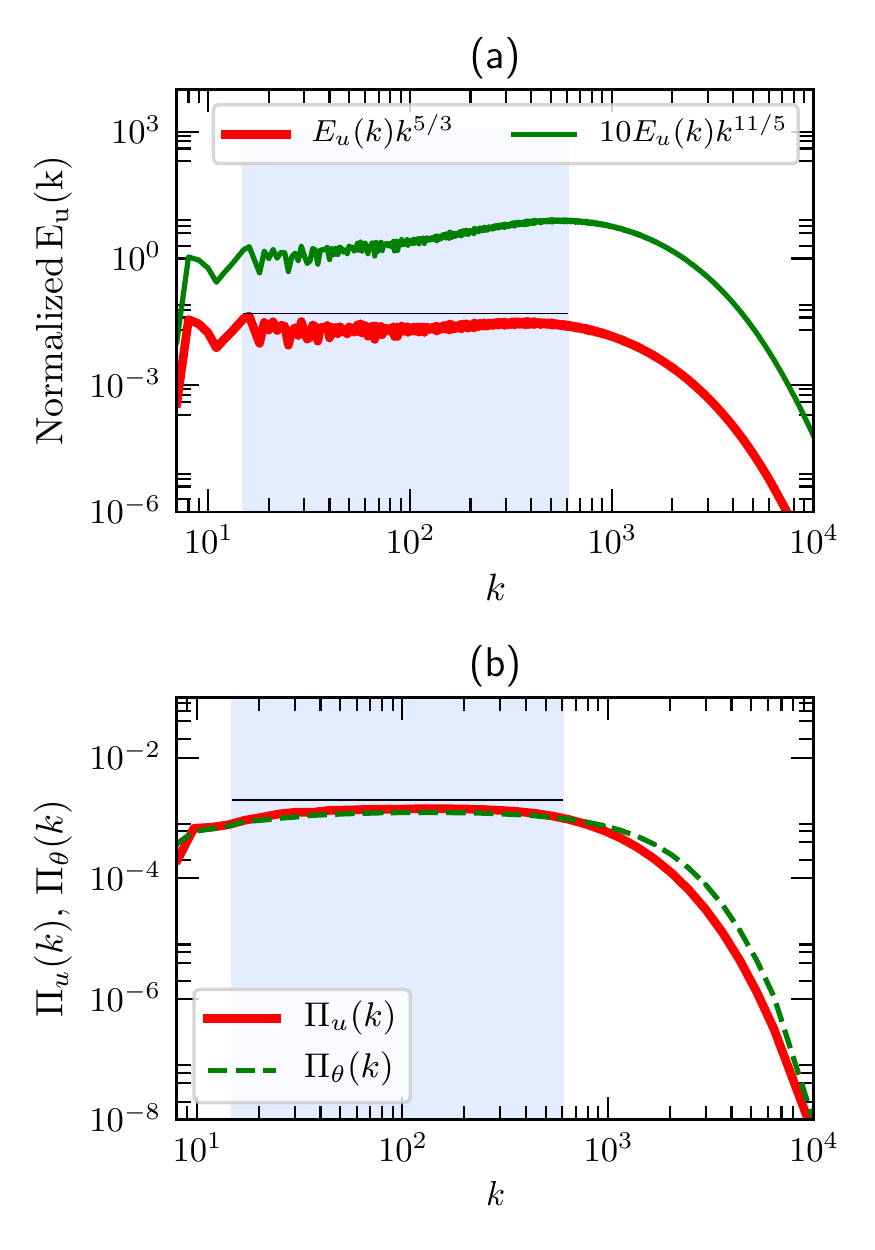}
\end{center}
\setlength{\abovecaptionskip}{0pt}
\caption{For RBC simulation on $4096^3$ grid with $\mathrm{Pr} = 1$ and $\mathrm{Ra} = 1.1 \times 10^{11}$: (a) Plots of the normalised kinetic energy  spectra $E(k) k^{5/3}$ (Kolmogorov-Obukhov, red curve) and $E(k) k^{11/5}$ (Bolgiano-Obukhov, green curve).  Flatness of $E(k) k^{5/3}$  indicates Kolmogorov-like phenomenology for RBC. (b) Plot of  kinetic energy  flux $\Pi_u(k)$ and entropy flux $\Pi_{\theta}(k)$, both exhibit a constant flux in the inertial range. From Verma {\em  et al.}~\cite{verma:NJP2017}. Reprinted under a CC BY licence ({http://creativecommons.org/licenses/by/3.0/}).}
\label{fig:RBC_spec}
\end{figure}

We use Tarang to study energy spectrum and energy flux of RBC and to resolve the long-standing question on the spectral indices.   Kumar {\em et al.}~\cite{Kumar:PRE2014} studied these quantities for $512^{3}$ grid and showed that the turbulent RBC exhibits Kolmogorov's spectrum similar to hydrodynamic  turbulence.  Here we present a more conclusive spectrum by performing RBC simulations on a $4096^3$ grid simulations using $65536$ cores with Rayleigh number $\mathrm{Ra} = 1.1 \times 10^{11}$ and the Prandtl number $\mathrm{Pr} = 1$. The energy spectrum and flux were computed using the steady-state data, which is achieved after around 1000 time steps. Fig.~\ref{fig:RBC_fig} exhibits  isocontours of constant temperatures exhibiting hot and cold structures.

In Fig.~\ref{fig:RBC_spec}(a) we exhibit the compensated energy spectra $E(k) k^{5/3}$ and $E(k) k^{11/5}$, which correspond to the compensated Kolmogorov spectrum and the Bolgiano-Obukhov spectrum respectively. The constancy of $E(k) k^{5/3}$ in the inertial range indicates that turbulent convection exhibits Kolmogorov spectrum, similar to hydrodynamic turbulence.   The energy flux, exhibited in Fig.~\ref{fig:RBC_spec}(b), is also constant in the inertial range, again confirming Kolmogorov-like phenomenology for RBC.  Thus our DNS helps resolve one of the outstanding questions in turbulent convection.  

\section{Conclusions}
\label{sec:conclusions}

In this paper  we perform scaling studies of a FFT library, FFTK, and a pseudospectral code Tarang on two different HPC supercomputers---Blue Gene/P (Shaheen I) and Cray XC40 (Shaheen II) of KAUST.  We vary grids from $768^3$ to $8192^3$ on cores ranging from 1024 to $196608$.  The number of  cores used for FFTK and Tarang are one of the largest in this area of research.  We also remark that on Blue Gene/P, the efficiency of FFTK is similar to that of  P3DFFT.   The main results presented in this paper are as follows:

\begin{enumerate}
\item We analyse the computation and communication times for FFTK.  We observe that the computation time $T_\mathrm{comp} \sim p^{-1}$ where $p$ is the number of cores, while the communication time $T_\mathrm{comm} \sim n^{-\gamma_2}$ where $n$ is the number of nodes.  

\item  Regarding FFTK, for Blue Gene/P, the communication time is comparable to the computation time due the slower core and faster switch. In Cray XC40 however the communication dominates  computation due to faster cores.   For FFTK, the total time scales as $T \sim p^{-\gamma}$ with $\gamma$ ranging from 0.76 to 0.96 for Blue Gene/P.  For Cray XC40,  $\gamma$ lies between 0.43 to 0.73.    In Sec.~\ref{subsec:FFT_Cray} we argue that the dragonfly topology of Cray XC40 is more efficient than torus topology of Blue Gene/P.  Yet, the speedup of interconnect for Cray XC40 is not much higher than that for Blue Gene/P.

\item Cray XC40 exhibits lower efficiency ($\sim 1.5\%$) than Blue Gene/P ($\sim 4\%$).  This is in-spite of the fact that ratio of the per-node compute power  of  Cray XC40 and Blue Gene/P is approximately 100.   The relative loss of  efficiency for XC40 is because the efficiency of its interconnect has not scaled in commensurate with that of its processor.   For communication intensive application like FFT,  the speed of interconnect is more important than that of the compute cores.  

The above observation indicates that  the performance of a HPC system depends on the application.  We need to be cautious about this, and factor into account the speed of the processor, interconnect, memory, and input-output.  In future, we plan to do an extensive study of these factors for FFT.

\item The fluid solver of Tarang exhibits weak and strong scaling on both the supercomputers.  The exponent $\gamma$ for Blue Gene/P varies from $0.8$ to $0.95$, but it ranges from $0.28$ to $0.68$ for Cray XC40.   

\item The solver for Rayleigh-B\'{e}nard convection also shows weak and strong scaling on both the supercomputers.  The corresponding $\gamma$ for Blue Gene/P ranges from 0.68 to 0.71, but it lies between 0.49 to 0.80 for Cray XC40.

\item The scaling of FFTK, and fluid and RBC solvers scale quite well up to 196608 cores of Cray XC40.  We however  observe saturation at 98000 cores for $768^3$ grid, possibly due to smaller data size. { To best of our knowledge, there is no such detailed scaling study on FFT  up to these many processors.}

\item The scaling of different basis functions (e.g.,  \texttt{FFF} and  \texttt{SFF}) are similar.  However the performance in the  \texttt{SFF} basis is slightly better than that in the  \texttt{FFF} basis.
\end{enumerate}

Thus, FFTK and Tarang scale nearly up to 196608 cores.  Thus these codes are capable of simulating turbulence at very high-resolution.  FFTK would also be useful for other applications, e.g., image processing, density functional theory, etc.

 \section*{Acknowledgement}
We thank Shashwat Bhattacharya for his  valuable help on plotting. Our numerical simulations were performed at Cray XC40 {\em Shaheen II} at KAUST supercomputing laboratory (KSL), Saudi Arabia. We thank KAUST scientists for the kind support while performing our simulations. This work was supported by research grants SERB/F/3279 from Science and Engineering Research Board India, K1052 from KAUST, and R\&D/20130307 from IIT Kanpur; and by KAUST baseline research funds of Ravi Samtaney.

\appendix
\section{Transpose-free Fast Fourier Transform}
\label{sec:appA}
{
In this appendix we describe how we avoid  local transpose in FFTK to save communication time.  For simplicity we illustrate this procedure using slab decomposition with complex data of size $n_0 \times n_1 \times (n_2/2+1)$.   The corresponding real space data is of the size $n_0 \times n_1 \times n_2$.

\begin{figure}[htbp]
    \centering
    \includegraphics[scale=0.18]{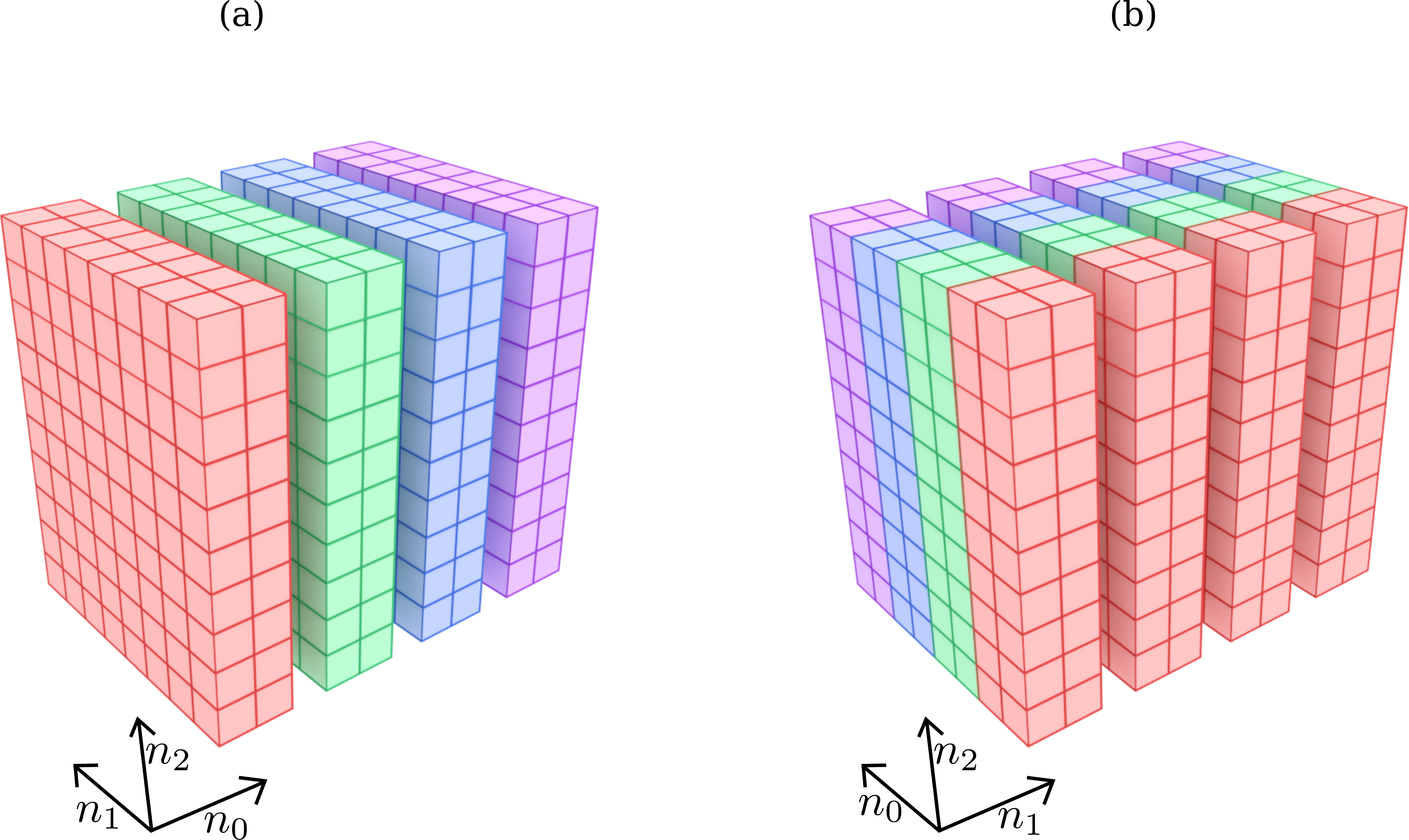}
    \caption{Data division for FFT with transpose: (a) Complex data of size $n_0 \times n_1 \times (n_2/2+1)$  in Fourier space. (b)  Real data of size $n_1 \times n_0 \times n_2$ in real space. Note that the axes $n_0$ and $n_1$ are exchanged during the transpose.  }
    \label{fig:3d_slab_transposed}
\end{figure}

The usual FFT implementation involving transpose is illustrated below.  The complex data is divided along $n_0$.  If there are $p$ processors, then each processor has $(n_0/p) \times n_1 \times (n_2/2+1)$ complex data. See Fig.~\ref{fig:3d_slab_transposed}(a) for an illustration.   A typical inverse transform  involves three steps:
\begin{enumerate}
	\item Perform two-dimensional inverse transforms (complex-to-real \texttt{c2r}) on $n_0/p$ planes each having data of size $n_1\times (n_2/2+1)$.
	\item Perform transpose on the array along $n_0$-$n_1$ axis.  This operation involves local transpose and  \texttt{MPI\_Alltoall} operations (to be described below).
	\item After the data transfer, the data along the $n_0$ axis resides in the respective processors.  Now in each processor, we perform one-dimensional real-to-real (\texttt{r2r}) inverse  transforms on $(n_1/p)\times n_2$ column each having data of size $n_0$.
\end{enumerate}
\begin{figure}[htbp]
    \centering
    \includegraphics[scale=0.7]{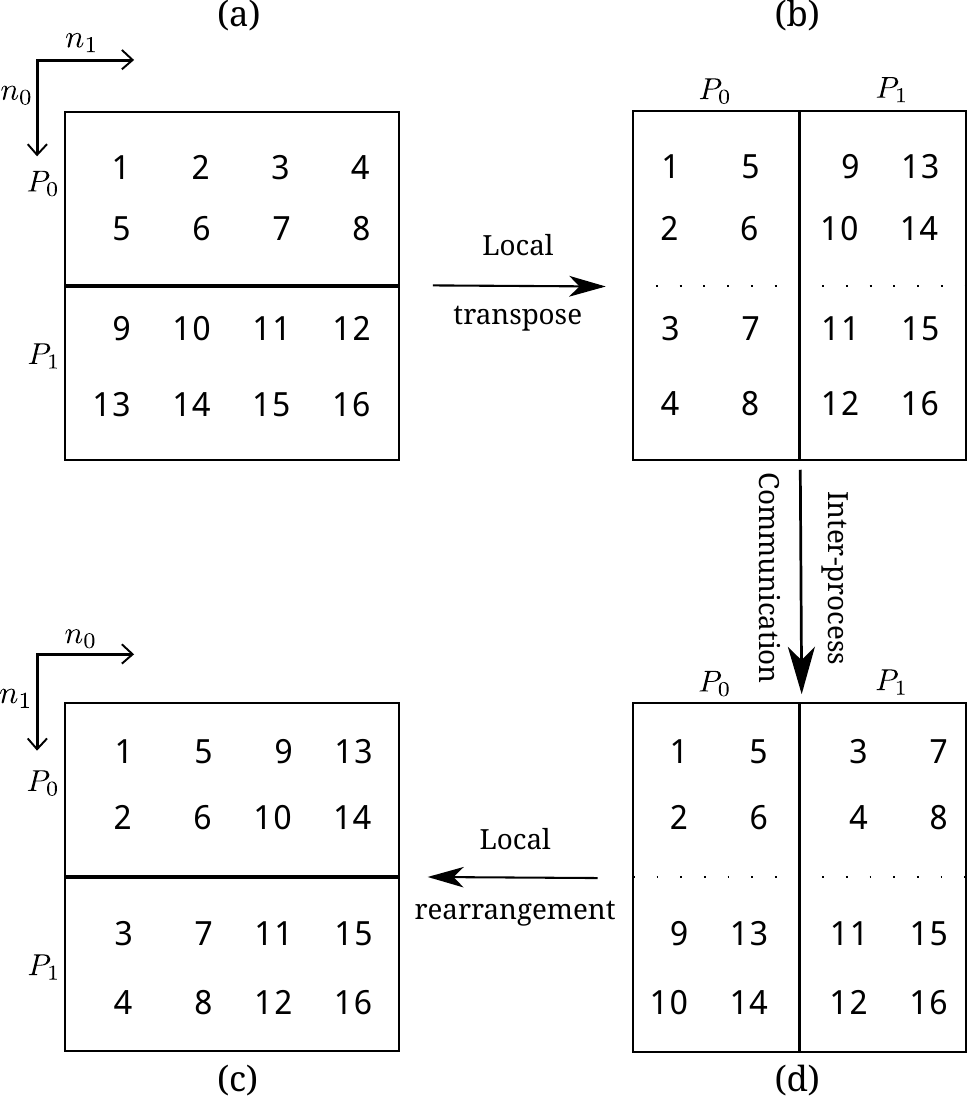}
    \caption{The standard transpose procedure in during a FFT.  It involves two local transposes and a \texttt{MPI\_Alltoall}.}
    \label{fig:transpose_standard}
\end{figure}
   In Fig.~\ref{fig:transpose_standard}, we illustrate this transpose operation using a simple example involving 16 data points and 2 processors.  In the first step, the local data is transposed, as illustrated in Fig.~\ref{fig:transpose_standard}(a,b).  In the example, in process $P_0$, the data [[1,2,3,4],[5,6,7,9]]  gets transformed to [[1,5],[2,6],[3,7],[4,8]].  After the local transpose, chunks of data are transferred among the processors using \texttt{MPI\_Alltoall} function (Fig.~\ref{fig:transpose_standard}(b) to Fig.~\ref{fig:transpose_standard}(c)). In this process, the blocks [[3,7],[4,8]] and    [[9,13],[10,14]] are exchanged between $P_0$ and $P_1$.  After this data transfer, there is another local transpose that transforms the data from Fig.~\ref{fig:transpose_standard}(c) to Fig.~\ref{fig:transpose_standard}(d).  
   
This complete operation is called  ``transpose" because it is similar to matrix transpose.  After transpose, we are ready for FFT operations along the $n_0$ axis.  Note that the data along the rows of Fig.~\ref{fig:transpose_standard}(d) are consecutive, that makes it convenient for the FFT operation.  We remark that the popular FFTW library employs the above procedure involving transpose.

An advantage of the above scheme  is that the FFT is performed on consecutive data sets that minimises cache misses.  However, the aforementioned FFT involves two local transposes, which  are quite expensive.   To avoid this, we have devised a FFT which is based on transpose-free  data transfer.  This process is described below.  

In the transpose-free procedure, we replace the transpose operations  (item 2 in the above list) with transpose-free inter-processor communication.   We employ  \texttt{MPI\_Type\_vector} and \texttt{MPI\_Type\_create\_resized} to select strided data to be exchanged among the processors. We illustrate the communication process in Fig.~\ref{fig:transpose}.    Here, the data block [[3, 4], [7, 8]] is transferred from $P_0$ to $P_1$, and the data block [[9, 10], [13, 14]] is transferred from  $P_1$ to $P_0$ using  MPI functions \texttt{MPI\_Isend/MPI\_Recv} or \texttt{MPI\_All\_to\_all}.  Note that  the data to be transferred are not consecutive, hence we need the MPI functions such as \texttt{MPI\_Type\_vector} and \texttt{MPI\_Type\_create\_resized} to create strided-data set. 
\begin{figure}[htbp]
    \centering
    \includegraphics[scale=0.7]{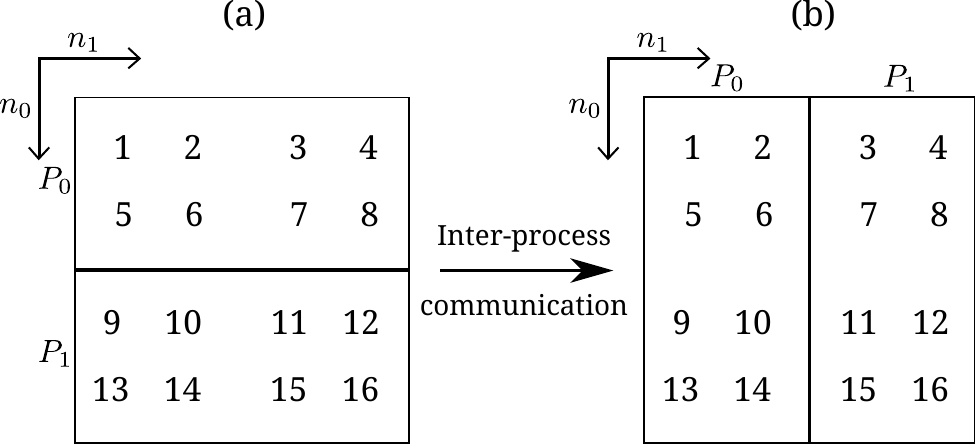}
    \caption{Transpose using strided \texttt{MPI\_Isend/MPI\_Recv} that does not require a local transpose. This is employed in the transpose-free FFT.}
    \label{fig:transpose}
\end{figure}
The data structure before and after the interprocess communication are shown in Fig.~\ref{fig:3d-slab}(a,b) respectively.  Here the data axes are not exchanged, however the columnar data  along  the $n_0$ axis are not contiguous. For example, the data of column  [1,5,9,13] of Fig.~\ref{fig:transpose}(b) are staggered by 1.  As a result, FFT along the $n_1$ axis involves consecutive data, but not along $n_0$.  The latter FFT however can be performed using  strided FFTW functions.  
\begin{figure}[htbp]
    \centering
    \includegraphics[scale=.2]{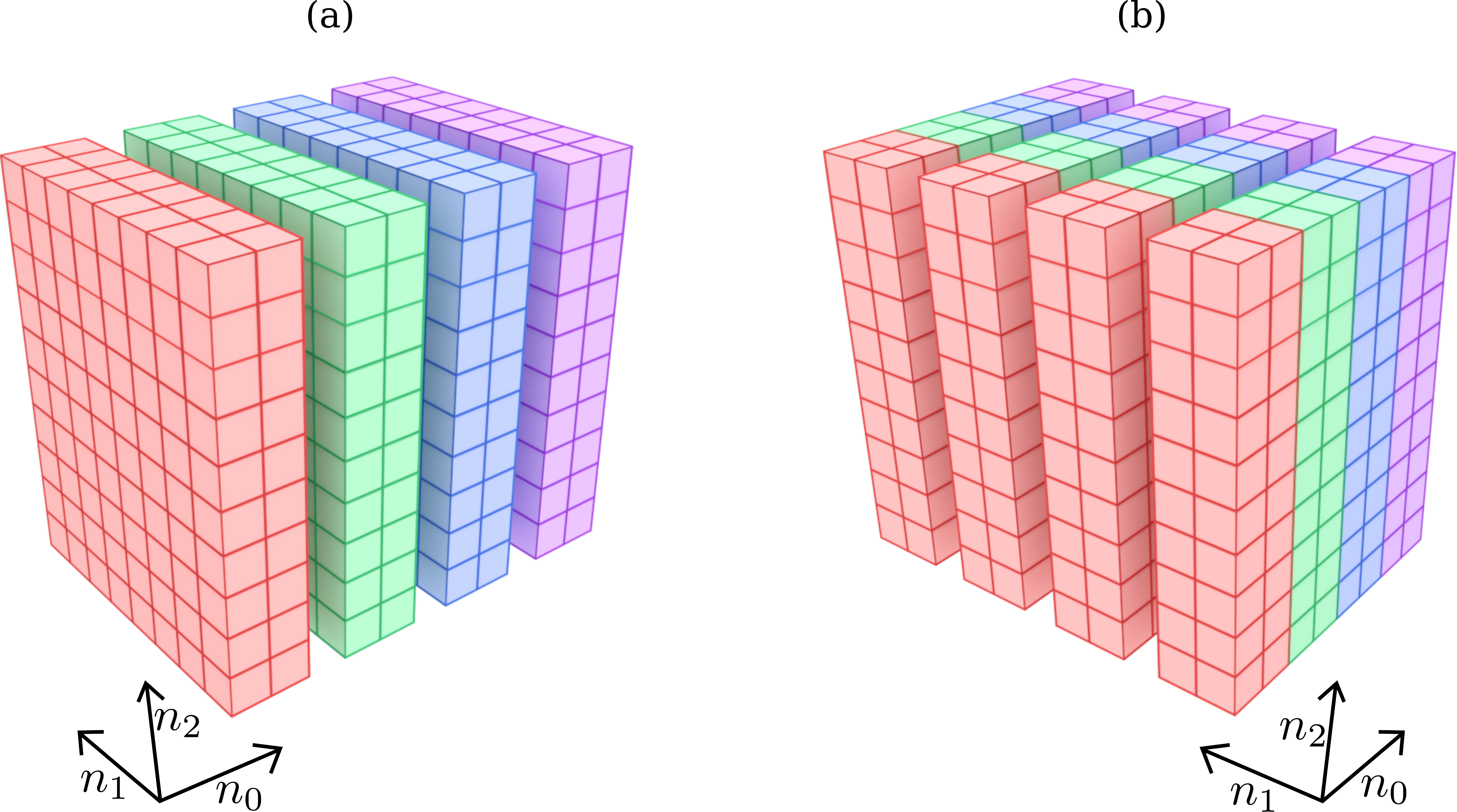}
    \caption{Data division for a transpose-free FFT: (a) Complex data of size $n_0 \times n_1 \times (n_2/2+1)$  in Fourier space. (b)  Real data of size $n_0 \times n_1 \times n_2$ in real space. Note that there is no exchange of  axes here.  Compare it with Fig.~\ref{fig:3d_slab_transposed}. }
    \label{fig:3d-slab}
\end{figure}

\begin{figure}[htbp]
    \centering
    \includegraphics[scale=0.44]{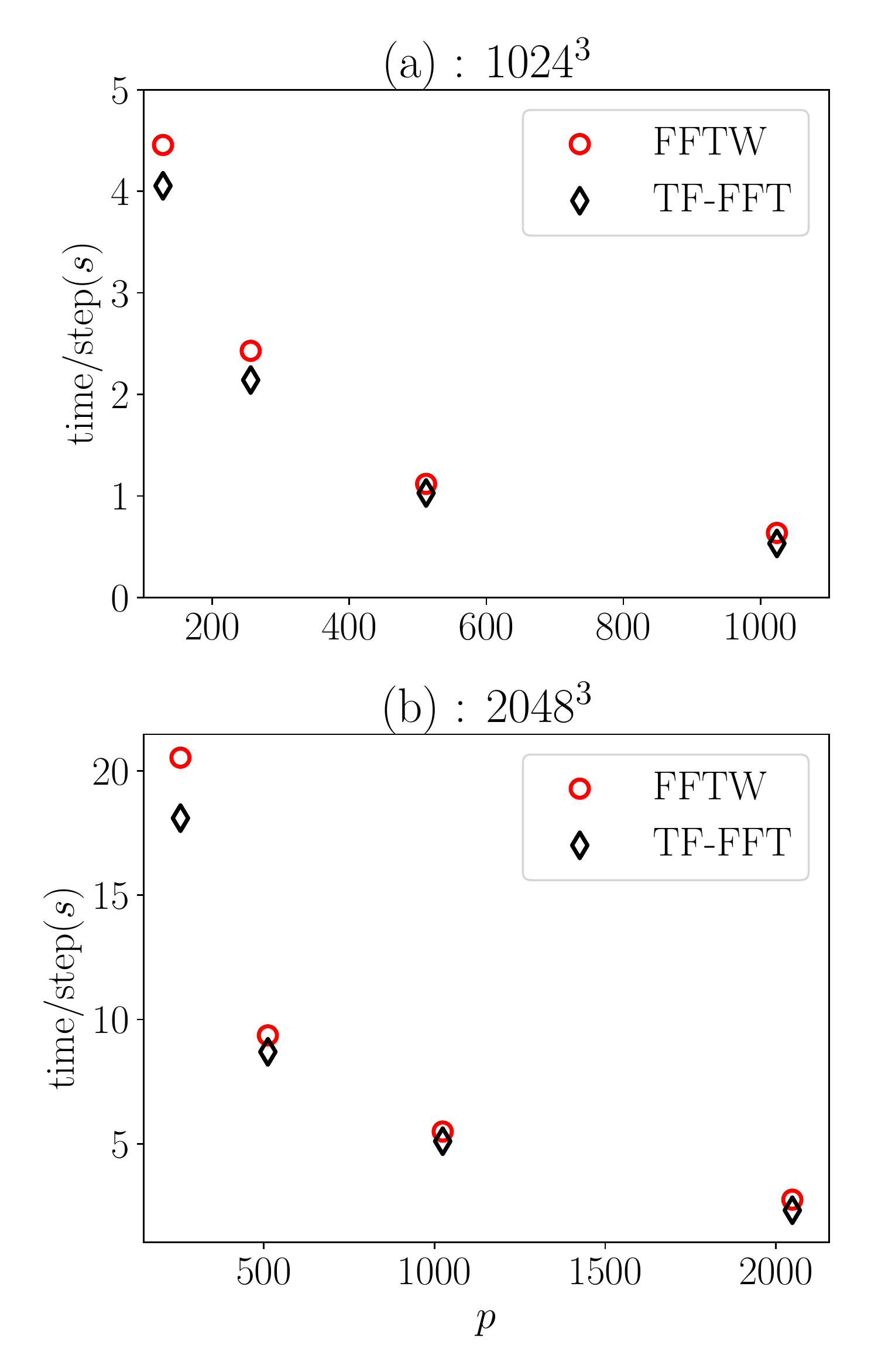}
    \caption{Comparison  between FFTs with transpose and without transpose for (a) $1024^3$ grid, and (b) $2048^3$ grid.  Transpose-free FFT is marginally superior than the one with transpose.}
    \label{fig:st-vs-fftw}
\end{figure}

Now let us briefly compare the performances of the two methods. The transpose-free scheme avoids  local transpose, hence it saves some communication time compared to the usual FFT.  A flip side of the transpose-free scheme  is that it needs  strided FFT that is prone to cache misses because the data are not contagious.  Note, however, that  intelligent cache prefetch algorithms~\cite{Berg:UW2004} could helps in efficient implementation of strided FFT.

To compare the efficiencies of the aforementioned FFT schemes, we performed FFTs using both the schemes.  Since FFTW involves local transposes, we use this as one of the benchmark programs.  We wrote a transpose-free FFT function as the other benchmark program. The tests were performed on  IBM BlueGene/P (Shaheen I) of KAUST for a pair of forward and inverse transforms on $1024^3$ and $2048^3$ grids. 

In Fig.~\ref{fig:st-vs-fftw}(a,b) we present the results for the $1024^3$ and $2048^3$ grids. In the figure the time taken by FFTW and transpose-free FFT are shown by red circle and black diamonds respectively.  We observe that the transpose-free FFT is  10\% to 16\% more efficient for $1024^3$ data, and 5\% to 14\% more efficient for $2048^3$ data.   The gain by the transpose-free FFT decreases as the number of processors are increased.  The difference in time is a sum of two factors: (a) gain by avoidance of local transpose, and (b) loss due to strided FFT.  We need a more detailed diagnostics to analyse the two algorithms.  For example, we need to separately compute the computation and communication time.  Also, it will be useful to estimate the time for the collection of the strided data, as well as that of the strided FFT.  These works will be performed in future.

The FFT operations in FFTK, which has pencil decomposition, in transpose-free. The only difference between the slab-based FFT described in this appendix and the pencil-based FFT is that the data exchange in pencil-based FFT takes place among the respective communicators. For example, among the \texttt{MPI\_COM\_ROW} and \texttt{MPI\_COM\_COL} communicators of Fig.~\ref{fig:pencil_combined}.
}


\end{document}